\documentclass{emulateapj}
\begin{document}

\title{Blazar 3C 454.3 in Outburst and Quiescence During 2005-2007: \\ Two Variable Synchrotron Emission Peaks}

\author{Patrick M. Ogle$^1$,  Ann E. Wehrle$^2$, Thomas Balonek$^3$, \& Mark A. Gurwell$^4$}

\affil{$^1$ Spitzer Science Center, California Institute of Technology, 
       Mail Code 220-6, Pasadena, CA 91125}

\affil{$^2$ Space Science Institute, 4750 Walnut Street, Suite 205, Boulder, CO 80301} 
\affil{$^3$ Department of Physics and Astronomy, Colgate University, Colgate, NY 13346}
\affil{$^4$ MS42, Smithsonian Astrophysical Observatory, 60 Garden Street, Cambridge, MA 02138}

\email{ogle@ipac.caltech.edu}

\shorttitle{Flaring Blazar 3C 454.3}
\shortauthors{Ogle et al.}
%\date{16 Mar 2010}

\begin{abstract}
We monitored the flaring blazar 3C 454.3 during 2005 June-July with the {\it Spitzer} Infrared 
Spectrograph (IRS: 15 epochs), Infrared Array Camera (IRAC: 12 epochs) and Multiband Imaging 
Photometer (MIPS: 2 epochs). We also made Spitzer IRS, IRAC, and MIPS observations from 2006 
December-2007 January when the source was in a low state, the latter simultaneous with a single 
{\it Chandra} X-ray observation. In addition, we present optical and sub-mm monitoring data. The 
2005-2007 period saw 3 major outbursts. We present evidence that the radio-optical SED actually consists of 
{\it two} variable synchrotron peaks, the primary at IR and the secondary at sub-mm wavelengths. The
lag between the optical and sub-mm outbursts may indicate that these two peaks arise from two distinct 
regions along the jet separated by a distance of 0.07-5 pc. 
The flux at 5-35 $\mu$m varied by a factor of 40 and the IR peak varied in frequency from 
$<1\times10^{13}$ Hz to $4\times10^{13}$ Hz between the highest and lowest states in 2005 and 2006, respectively. 
Variability was well correlated across the mid-IR band, with no measurable lag. Flares that doubled in flux
occurred on a time scale of $\sim 3$ days. The IR SED peak moved to higher frequency as a flare 
brightened, then returned to lower frequency as it decayed.  The fractional variability amplitude 
increased with frequency, which we attribute to decreasing synchrotron-self absorption optical depth. 
Mid-IR flares may signal the re-energization of a shock that runs into inhomogeneities along the 
pre-existing jet or in the external medium. The synchrotron peak frequencies during each major outburst
may depend upon both the distance from the jet apex and the physical conditions in the shocks. Variation
of the Doppler parameter along a curved or helical jet is another possibility. Frequency variability of the IR
synchrotron peak may have important consequences for the interpretation of the blazar sequence, and the presence 
of a secondary peak may give insight into jet structure.

\end{abstract}

\section{Introduction}

In 2004 August - 2005 September, the blazar 3C 454.3 (z=0.859) underwent the largest radio-optical 
flare in its recorded history. In 2005 May, it briefly surpassed 3C 273 as the optically brightest quasar 
in the sky in spite of its much greater distance. This flaring event was intensively monitored 
at all frequencies by observers all over the world, using both ground and space-based 
observatories \citep{vrb06,pfb06,fcm06,vra07,rvl07}. The source has since flared twice again with smaller 
amplitude, offering continuing opportunities to study this spectacular phenomenon, e.g., 
\cite{vil09,hagen09, ver09,Rai08,Rai08b}.

Blazars, including flat-spectrum radio-loud quasars and BL Lac objects, are characterized 
by strong variability and high polarization. According to active galaxy unification models, 
they are viewed within $\sim 20\arcdeg$ of the radio jet axis. The spectral energy 
distribution (SED) is dominated by relativistically boosted emission from the core of the jet. 
The core is strongly variable at all frequencies, owing to rapid movement and changes 
in the jet on sub-parsec scales. Very-long baseline interferometry typically shows 
jet components with apparently superluminal motion at milliarcsecond resolution, 
demonstrating the importance of relativistic effects.

The SEDs of blazars are typically characterized by two major bumps, one which peaks in the radio 
through X-ray bands, and the other which peaks in X-$\gamma$ ray bands. The low frequency 
bump is highly polarized, indicating synchrotron emission from relativistic electrons in a 
magnetized jet. The high frequency bump is thought to be produced by inverse-Compton
scattering of photons by relativistic electrons. However, the source of the photons is a subject 
of debate, ranging from the synchrotron photons in the jet itself to external photon fields 
\citep[e.g.,][]{w98}. In some blazars, a significant fraction of the optical-UV thermal emission 
as well as broad optical-UV emission lines may arise from an accretion disk \citep{seb88}. See, for 
example, \cite{ghis09} and references therein, for a review of models and their application to the 
class of flat spectrum radio quasars.

Theoretical unifying schemes for $\gamma$-ray bright blazars have been proposed: the
more luminous the blazar, the lower the peak frequencies of the synchrotron and Compton
bumps \citep{fmc98, gcf98}. Physically, the down-shift in peak frequency can be caused
by increased Compton cooling at higher radiation densities. According to the 
external-Compton (EC) model, photon fields from the accretion disk, broad-line
region (BLR), and dusty torus act to sap energy from the relativistic jet. Hence, luminous 
flat-spectrum radio quasar (FSRQ) SEDs peak (on average) at lower frequencies than do those 
of BL Lac objects. Similarly, we might  expect accretion disk luminosity variations in a 
single object to be accompanied by changes in the frequency of the synchrotron peak. 

However, Compton cooling may be counterbalanced by re-acceleration of high energy electrons
in jet shocks. As a disturbance travels down the jet, it may run into inhomogeneities
internal to or external to the jet. At such locations, the bulk relativistic kinetic
energy and energy in the electromagnetic field is dissipated, heating the electrons, which 
emit copious synchrotron photons as they gyrate about the magnetic field lines. The increased 
number of high energy electrons will shift the synchrotron peak to higher frequencies. Thus 
the peak synchrotron frequency is determined by a balance between shock heating and Compton 
cooling, and should vary as jet shock components evolve. An exact correspondence 
between accretion luminosity and peak synchrotron frequency is therefore not expected.

The IR is a crucial band since it covers the primary synchrotron emission peak of FSRQs,
which may provide seed photons for the Compton $\gamma$-ray bump. In addition, while the base 
of the jet is optically thick at radio frequencies, it is optically thin at mid-IR through 
UV frequencies. Thus IR observations are less subject to optical depth effects than radio 
observations and can be used to probe short variability time scales which arise close to the 
origin of the jet. If flaring episodes are caused by shocks \citep{mg85}, then flares should 
occur simultaneously across the IR through UV wavebands \citep{l94,swv02}. Therefore, IR 
spectral variability should reflect changes in the energy distribution of relativistic electrons. 
Variability on day to month time scales can be used to study the evolution of the shock as 
it propagates along the jet and through the environment of the active galactic nucleus. 

The blazar 3C 454.3 has undergone a number of flaring episodes, with bright radio 
outbursts occurring roughly once every 6 years \citep{cba04}. It has a one-sided 
milliarcsecond scale radio jet \citep{prw80, cg96, mjm02, sfh04}. Very Long Baseline 
Interferometric (VLBI) monitoring observations reveal a series of jet components, 
including stationary components and components with apparent superluminal motion of 
up to 16 $h^{-1} c$ \citep{pt87,gm99,jmm01, klh04}.

Historically, the synchrotron emission from 3C 454.3 typically peaks in $\nu F_\nu$ in 
the far-IR  ($\sim 100$ $\mu$m). IR emission observed by ISO had a flux of 34 mJy at 
12.8 $\mu$m \citep{hmb04}. Excess emission in the SED at 60 $\mu$m was attributed to thermal 
emission from 100 K dust by these authors, but we will present evidence that
this emission is highly variable.  The X-ray spectrum observed with {\it Beppo-SAX} was 
characterized by an 0.5-200 keV power law with spectral index $\alpha=-2.3$ \citep{tmg02}.The greatest 
amount of energy is emitted in the Compton bump, which peaks at $\sim 10$ MeV.  Gamma-ray emission was 
observed at energies of 0.05-4 GeV by the EGRET and OSSE instruments on the {\it Compton} Gamma-Ray 
Observatory \citep{hbd93,mnb95,zcs05}.  Gamma-ray detections after the data in this paper were obtained 
include those by AGILE \citep{ver09} in 2007, by the LAT instrument on Fermi in 2008 \citep{abdo09}, and 
by AGILE in 2008-2009 \citep{donna09}; see also the review by \cite{ver09b}.  

We were awarded director's discretionary time to observe 3C 454.3 daily with the {\it Spitzer} 
Infrared Spectrograph (IRS), the Infrared Array Camera (IRAC), and the Multiband Imaging Photometer 
(MIPS) over a period of 4 weeks in 2005 July, during normally scheduled instrument campaigns. We also 
took coordinated {\it Spitzer} MIPS, IRS, IRAC and {\it Chandra} High Energy Transmission Grating
(HETG) observations in 2006 December -2007 January.  We obtained observations from  long term optical 
monitoring programs at Colgate University and Palomar Observatory and from calibrator monitoring at 
the Submillimeter Array (SMA). In this paper, we present these data and discuss variability and the 
nature of synchrotron emission flares from the blazar 3C 454.3. 

\section{A Brief History of 3C 454.3 in 2005-2007}

Observations that were made during the same time period as those reported in this paper included several 
large multiwavelength campaigns, summarized here. Following the alert of an optical flare in 3C 454.3 by 
\cite{bal05a} and \cite{bal05b}, \cite{fcm06} obtained near-IR/optical photometry starting two days later. 
\cite{vrb06}  made the first report of the WEBT multiwavelength campaigns, including ToO pointings by Chandra 
and Integral, that followed the discovery of the May 2005 optical flare. \cite{rvl07} reported the detection 
of the "small and big blue bumps" during the low state in 2006. 
\cite{vra07} derived the delay between centimeter-band radio and optical light curves during 2005-2006. 
\cite{gio06} observed 3C 454.3 with the robotic Rapid Eye Mount optical/near-infrared telescope and with the 
{\it Swift} satellite in April-May 2005.  \cite{Rai08} presented multifrequency observations by the WEBT and 
XMM-Newton in 2007-2008, overlapping with {\it AGILE}'s November 2007 observations.  They also used 1.3 mm 
monitoring data from the SMA, some of which is also used in our paper, over the time period 2005-2008 (see 
their Figure 4) as well as 1.3 mm data from Pico Veleta, Spain. X-ray data from  {\it Swift}, {\it Chandra}, 
{\it XMM-Newton}, and {\it Integral} data from the 2005-2007 period have been presented by \cite{gio06}, 
\cite{vrb06}, \cite{rvl07} and \cite{pfb06}. We describe the results of these papers as follows.

1. \cite{fcm06} observed 3C 454.3 with the Automatic Imaging Telescope of Perugia University Observatory and 
the Rapid Eye Mount telescope in Chile, obtaining V,R,I and H band photometry. The spectral index over these 
bands showed no strong significant changes. 

2. \cite{vrb06} presented results up through September 2005, finding that the source was redder when brighter. 
 A mm outburst occurred in June-July 2005, followed months later by the 37-43 GHz peak. Chandra and Integral 
X-ray observations in May 2005 showed unusually high fluxes.

3. \cite{gio06} observed 3C 454.3 with the robotic Rapid Eye Mount optical/near-infrared telescope and with the 
{\it Swift} satellite in April-May 2005. They found that the optical and ultraviolet flux doubled within a 
single $1.5 \times 10^{5}$ second exposure, the  XRT (2-10 keV) X-ray flux varied little during the same time. 
However, on time scales of a few days, the BAT (15-150 keV) X-ray flux varied by more than a factor of three; 
in contrast, the average level of the optical-ultraviolet flux was approximately constant between the two UVOT 
observations of  24 April and 17 May 2005.

4. \cite{vra07} reported multiwavelength observations during 2005-2006. Their analysis suggested that the big 
radio flare (43-37 GHz) in early 2006 was associated with a minor optical flare in October-November 2005, not 
with the spring 2005 major optical flare.  A combination of disturbances traveling down the jet and changes of 
viewing angles of different emitting regions, with concomitant changes in Doppler boosting, were found to 
explain the radio delays with respect to the optical emission.

5. \cite{rvl07} detected the "small and big blue bumps" during the low state in 2006.
Observations from K-band through ultraviolet  showed a small peak in optical band (V), an upturn 
in ultraviolet from U-band to UVW1 and UVM2 and an upturn at the infrared end. The infrared end was probably 
variable synchrotron emission, complicated by a non-variable H-alpha line (reported by \cite{Rai08}
via spectroscopy in J band). The "small blue bump" is probably a blend of iron lines, Mg II lines and Balmer
 continuum from the broad line region. The ultraviolet upturn, interpreted as the beginning of the big blue 
bump was probably from thermal emission of the accretion disk.  All these physical emission mechanisms 
contributed during various activity levels. The underlying accretion disk and broad line contributions were 
visible most clearly when the synchrotron emission was at low ebb. The X-ray spectra could be fitted with a power 
law, but seemed to require extra $N_{H}$ absorption some of the time. Alternatively, there may have been spectral 
changes in the soft X-ray range. 

6. \cite{Rai08} presented multifrequency observations by the WEBT and XMM-Newton in 2007-2008, overlapping 
with AGILE's November 2007 observations.  They also used 1.3 mm monitoring data from the SMA, some of which 
is also used in our paper, over the time period 2005-2008 (see their Figure 4) as well as 1.3 mm data from 
Pico Veleta, Spain. Correlation of the WEBT R-band optical data with the SMA and Pico Veleta data showed that 
the millimeter fluxes lagged the optical fluxes by 40-80 days, with Discrete Correlation Function maximum signal 
at 65 days. Taking only non-outburst data for the correlation yielded a DCF maximum at 20 days.  They suggested 
that the ``jet regions emitting the optical and mm radiations are now better aligned than in the past, and/or 
that the opacity in the jet has decreased, allowing the release of mm radiation closer to the optically emitting 
zone.''  We return to this correlation later in our paper.  Several multiwavelength SEDs during  2005-2007 are 
shown and analyzed.  From the X-ray-optical correlation, they found that the X-ray, near-infrared and optical 
emission would be produced in the same spatial region.

\section{Observations}

\subsection{Submillimeter Array Observations}

The Submillimeter Array (SMA) is an 8-element radio interferometer located atop Mauna Kea, Hawaii, 
which operates in the 1.3 mm, 850 $\mu$m, and 450 $\mu$m atmospheric windows \citep{ho04}.  
In the 1.3 mm and 850 $\mu$m windows, quasars are utilized as amplitude and phase gain calibrator sources 
during routine observing, and the flux density scale is derived through observations of standards in each 
session, typically solar system sources (with Uranus, Neptune, Titan, Ganymede and Callisto being the most 
used).  The SMA supports a program to calibrate quasar flux densities using both routine science and 
dedicated calibration observations, and provide their histories to SMA users and the wider astronomical 
community.  The Submillimeter Calibrator List, containing the flux measurement histories of several hundred 
quasars, can be found in the Tools section of the SMA Observer Center \footnote{http://sma1.sma.hawaii.edu}, with 
further details provided by \cite{gu07}. Data for 2005-2007 are presented in {\bf Table} 1  and 
{\bf Figure} 1 for the 850 $\mu$m and 1 mm bands.  

\subsubsection{Submillimeter Flares}

The sub-mm data, shown in {\bf Figure} 1a, show a huge flare peaking at 42.7 Jy on 2005 June 24 and 
a smaller flare of 20 Jy in February 2006. {\bf Figure} 1b shows the extraordinary flare in 2005 and a surprisingly 
large amount of substructure during the 2-month period around the peak, including five dates on which the flux 
exceeded 40 Jy.  We think the substructure is real rather than random noise because the variability  correlates 
well between the 230 and 345 GHz band measurements with several measurements in a row  showing consistent increasing 
or decreasing flux density.  We can not completely exclude systematic error, for example, observer bias in knowing 
what the preceding measurement value was. 

The SMA is sensitive to only a single polarization which also rotates on the sky as a function of elevation. Anecdotal 
reports by other observers indicated that 3C 454.3 was not polarized more than a couple of percent. Given the complex 
rotation relative to the source, it is unlikely that any significant structure in the time variability would be 
related to changes in the polarization angle, but it cannot be excluded without considerable extra work beyond the scope 
of this paper.

For about 98\% of the measurements, the frequency quoted is in fact the local oscillator (LO) frequency, and 
the flux measurement is the {\it average} of the lower and upper sideband values (in the heterodyne mixing, 
the sky frequencies are mixed down to the more manageable IF frequency range of 4-6 GHz, but we accept both 
sidebands, e.g. LO +/- (4-6 GHz).  The use of interferometry and the complex correlator allows us to unambiguously 
separate the sidebands, so we really get two flux measurements, separated by 10 GHz. The SMA 3C 454.3 data yield a 
spectral index that typically lies near $\nu^{-0.5}$, thus we would expect only a 2\% drop in flux from 220 to 
230 GHz (1.3 mm band), and only 1.5\% from 335 to 345 GHz (850 $\mu$m band).  The peak substructure in the sub-mm flare 
is definitely not caused simply by changes in the frequency of observation.

\subsection{Spitzer Observations}

We observed 3C 454.3 with {\it Spitzer} IRS, IRAC, and MIPS at several epochs during 2005-2007 
({\bf Tables} 2 and 3). Note that the three instruments were operated separately (according to their regularly
scheduled campaigns of typical length 1-2 weeks) so we do not have simultaneous coverage in
the respective IR bands. First, we observed the source once per day from 2005 June 30 - 2005 July 27 
to track mid-IR spectral variability following the 2005 May outburst. We conducted  a series of 14 daily 
IRS observations between 2005 June 30 and July 12, followed by 12 daily IRAC observations between July 14 and 
July 26, followed by two MIPS observations on July 27, the last day of the Spitzer visibility window.
We observed the source again on 2006 Dec 20 - 2007 Jan 1, the first of two coordinated {\it Spitzer} and 
{\it Chandra} observation sequences (the second is reported by Wehrle et al. 2010, in preparation). We made an 
observation with IRS on 2006 Dec 20, with IRAC on 2006 Dec 25 and with MIPS on 2007 January 01, the latter 
simultaneously with Chandra as part of a program unrelated to the flare.  All IRS observations were made in 
staring mode, and IRAC and MIPS images were taken in mapping modes. Parameters of the observations are given in 
{\bf Tables 2 and 3}.

\subsubsection{IRAC}

The IRAC images ({\bf Fig. 2}) on 2005 July 14 (bright state) and 2006 Dec 25 (faint state) show the unresolved 
quasar in a field of stars and foreground galaxies.  The Spitzer Science Center (SSC) pipeline mosaic data 
(version 14.0.0) were used for aperture photometry carried out with SSC APEX software package in beta release which 
was the only Spitzer software available for photometry when we began this work.  The apertures used were 2, 4, 5, and 5 
pixels in radius (1 pixel = 1.2 arcsecond) at 3.6, 4.5, 5.8 and 8.0 $\mu$m, and corrected using the values given in 
{\bf Table} 5.7, p. 53 of the IRAC Data Handbook, Version 3.0. We cross-checked the computed quasar fluxes for a range of 
apertures with those of a nearby star of roughly comparable brightness; the errors are consistent with the 1.5\% systematic 
errors plus 3\% absolute calibration uncertainty of the IRAC \citep{rmc05}. Despite the unpredictable brightness of the 
source, none of the IRAC images were saturated. Following the release of a new version of IPAC's Skyview software (courtesy 
of B. Hartley (IPAC)), we used it to reduce the images independently. The APEX and Skyview results agree to within two 
percent, except on 2005 July 14, where the APEX value exceeds the Skyview value by 10\% at 3.6 $\mu$m.  The Skyview 
fluxes at 3.6, 4.5, 5.8, and 8 $\mu$m are listed in {\bf Table 4}. 

Serendipitously, during the faint state, the 3.6 and 4.5 $\mu$m IRAC images displayed a faint, 10-$\sigma$ feature about 6 "
from the quasar at PA $\sim 150$ degrees. IRAC detections have been made of arcsecond-scale jets from other blazars, 
however, this feature appeared on the {\it opposite} side from the 8" radio jet at PA -45 degrees \citep[e.g.,][]{Cooper2007}. 
The feature was present in all individual frames when the source was faint, including those from another of our Spitzer 
programs. It was not associated with typical IRAC instrumental artifacts (J. Surace, private communication).  Working with 
Marco Chiaberge, we pulled archival ACS Hubble images (F. Tavecchio, PI) of the source during a faint episode in August 2004, 
and found a plausible identification of the IRAC feature with two galaxies, conflated by the much larger IRAC PSF. The 
galaxies' brightness and colors indicated that they could be located at roughly the same distance as 3C 454.3 (W.C. Keel, 
private communication). {\bf Figure 3} shows the IRAC and Hubble images side by side.

\subsubsection{MIPS}

The MIPS images ({\bf Fig. 4}), taken when the source was bright on 2005 July 27 at 01:48 UT  
and 15:11 UT, and when the source was faint on 2007 January 1, show the unresolved quasar. In the 
full 24 $\mu$m images, the bright quasar and a few faint galaxies were visible. At 70 $\mu$m, the quasar was 
isolated on fairly smooth background. The first Airy ring is clearly visible in each image.  In the 2005 
July 27 datasets, evidence of "soft saturation" was found in the 160 $\mu$m data (the pipeline images showed a 
doughnut-like structure instead of a smooth peak).  Special MIPS custom processing on the BCD-level data was performed 
to obtain useful fluxes  by MIPS Instrument Team member K. Gordon (see \cite{g05} for software description). For 
24 and 70 $\mu$m data, the pipeline mosaic data (version 13.2.0) was adequate for our use. The 2007 Jan 01 data 
were not affected by soft saturation; we used pipeline data (version 13.2.0). The IPAC Skyview analysis  package  was 
used for aperture photometry.  At 24 $\mu$m, we used a 13" radius aperture and applied an aperture correction of 1.164 
(Table 3.13, p. 33 of MIPS Data Handbook version 3.2). At 70 $\mu$m, we used 35" radius aperture and applied an 
aperture correction for a $\nu^{-2} $ spectral index of 1.197 (Table 3.14, p. 33, MIPS Data Handbook version 3.2).  
For our 160 $\mu$m data on 2007 Jan 1, we used 50" radius aperture and applied an aperture correction of 1.470 (Table 
3.16, p. 35, MIPS Data Handbook v. 3.2). 

The fluxes on 2005 July 27 at 01:48 UT and 15:11 UT at 24 $\mu$m  were $1.12$ and $1.16$ Jy, at 70 $\mu$m, both epochs' 
fluxes were $3.51$ Jy and at 160 $\mu$m, $5.43$ Jy, with systematic errors of 10\%, 10\% and 20\% respectively. The 
fluxes on 2007 Jan 01 at 24, 70 and 160 $\mu$m  were $0.060$ Jy and $0.189$ Jy and $0.22 $ Jy, with 
systematic errors of 10\%, 10\% and 30\% respectively.

Additional confidence in the Spitzer low state 160 micron flux density of 0.22 Jy with 30\% systematic uncertainty, measured 
on 2007 Jan 1, is provided by the completely independent Infrared Space Observatory Photometer measurements while 
the quasar was in an earlier low state: ISO-Phot measured the 120 and 180 micron flux densities of 0.187 and 0.248 Jy on 
1997 December 18 \citep{hmb04}.  ISO-PHOT and Spitzer MIPS were calibrated using different primary and secondary
calibrators, including asteroids.  The 3C454.3 maps made by ISO and Spitzer were in different orientations on the sky.

\subsubsection{IRS Low Resolution}

IRS was used in standard staring mode to take low-resolution spectra with the Short-Low (SL2,1) and 
Long-Low (LL2,1) modules ({\bf Figs. 5-6}). Because of the unexpectedly bright flux level, digital 
saturation occurred in some of the 2005 July exposure ramps. This effect is automatically corrected 
for in the data reduction pipeline by throwing out the saturated samples, leading to lower effective 
exposure time for the brightest pixels and a corresponding reduction in S/N. Several pixels in the LL1 
order of the epoch 14 spectrum had only 2 unsaturated samples in each ramp, leading to the noise spike 
at 22 $\mu$m ({\bf Fig. 5}). To guard against this problem, we reduced the exposure times and increased 
the number of exposures in subsequent epochs.

Spectral reductions began from the Basic Calibrated Datasets (BCDs), which
have been processed using the {\it Spitzer} S15.3.0 pipeline. The pipeline
applied subtraction of dark current, ramp fitting, and nonlinearity, droop, and stray-light
corrections. The 2D BCD spectra were median-averaged at each of the two standard nod 
positions and off-source sky background was subtracted. Radiation-damaged "rogue" pixels were 
interactively selected and median-filtered from the 2D spectra using the IRSCLEAN program. 
Spectra for each nod position were extracted using standard tapered windows (SL2: $7\farcs2$ 
at 6 $\mu$m, SL1: $14\farcs 4$ at 12 $\mu$m, LL2: $20\farcs4$ at 16 $\mu$m, LL1: $34\farcs4$ 
at 27 $\mu$m). Finally, spectra from the two nod positions were averaged. 

We have devised a fringe-correction algorithm to ameliorate spectral fringes in the
LL spectra. The wavelengths of the detector fringes depend on where exactly the source is placed
in the slit, which varies with the pointing error of the telescope. The detector fringe
patterns were extracted from flat field images and used to derive fringe correction
functions. The fringe spectrum was shifted and divided by itself to generate a correction
curve to minimize the residual fringes in the source spectrum. This procedure reduced 
the residual fringe amplitudes from $<4\%$ to $<2\%$ in LL.

We found significant offsets between low-resolution spectral orders, which were likely caused by 
random pointing offsets. This effect was most pronounced in the SL1 order because of the small size 
of the slit relative to the telescope point-spread function. Multiplicative corrections of 1.0-1.2 
(constant with wavelength) were applied to match the flux in each of the SL2, SL1, and LL2 spectral 
orders to LL1 to within an accuracy of 1\%. The corrections were largest in the epoch 6-8 SL1 spectra 
(8-20\%). For these epochs, there was a residual convex downward curvature of $\sim 3\%$ in SL1 caused 
by wavelength dependent slit losses, which we did not correct, but which do not
affect our conclusions.

We used flux calibrations provided by the SSC, employing low-order 
polynomial fits to standard star observations to convert from electron s$^{-1}$ to Jy.
The absolute flux calibration accuracy is limited by the uncertainty of the stellar
atmosphere model for the flux calibration standard HR 7341, which is estimated to be $3-5\%$ across the 
IRS wavelength band \citep{de07,dma04}. The relative flux accuracy of each epoch is
limited by slit losses. Repeatability observations of the standard star HD 173511 give 
independent 1-$\sigma$ flux calibration uncertainties of 2\% for SL2, 1\% for SL1 and LL2, 
and 3\% for LL1. After matching all orders to LL1, the relative flux accuracy is 3\%. 
We achieve S/N values of $\sim 60-120$ in the low-resolution spectra, limited by remaining 
systematic uncertainties in the flat field at a level of $1-2\%$.

We measure the mean flux in five wavebands ({\bf Table 5}: 5.5-6.5,11.5-12.5, 17-19, 23-25, and 28-32 $\mu$m, observed) 
from the IRS low resolution spectra. The mean wavelengths for these bands are centered 
at 6.0, 12.0, 18.0, 24.0, and 30.0 $\mu$m. The light curves for these wavebands are presented in Section 3.2.

There are no emission lines or absorption lines stronger than 2\% of the continuum flux in 
the 2005 June-July low resolution spectra ({\bf Fig. 5}). There is also no evidence for any significant 
broad silicate absorption or emission or any emission from polycyclic aromatic hydrocarbons (PAHs). 
Small deviations of $< 2\%$ come from uncertainties in the flux calibration introduced 
by residual detector fringing effects. The most prominent instrumental feature is a $\sim 
10\%$ bump at 14 $\mu$m (the so-called SL1 ``teardrop''), which may be caused by scattered light 
in the detector. There is a similar instrumental bump at 7 $\mu$m and a dip at $\sim 5.5$ $\mu$m,
both in SL2 and of unknown origin.

\subsubsection{IRS High Resolution}

We also took high resolution spectra ({\bf Fig. 7}) with the Short-High (SH) and Long-High
(LH) modules at the beginning and end of the 2005 IRS campaign (June 30 and July 13).
These observations were designed to search for any narrow emission or
absorption features in the mid-IR spectrum.  The high resolution spectra are divided 
by a factor of 1.04 to match the low-resolution flux calibration, since there is a level 
mismatch between the two flux standards used in the S15 pipeline to calibrate low and 
high-resolution data \footnote{http://ssc.spitzer.caltech.edu/irs/features.html}. No sky 
subtraction was performed. The sky background was small compared to the source flux in SH for 
the brighter of the two epochs. However, it became significant for the larger LH slit and for 
the fainter epoch.

We subtracted a second-order polynomial fit to each order of the high-resolution spectra in order to 
get a closer look at any possible emission line features ({\bf Fig. 8}). 
On inspection, it is clear that there is no significant detection at either epoch of any of the following 
quasar or host galaxy emission features at a redshift of $z=0.859$: [Ar {\sc ii}], H$_2$ S(3), [S {\sc iv}], 
11.3 $\mu$m PAH, [Ne {\sc ii}], [Ne {\sc v}], [Ne {\sc iii}], or [S {\sc iii}]. There is a possible detection 
of [Ne {\sc vi}] $\lambda$7.65 $\mu$m at 14.22 $\mu$m. However, it is most likely caused an 
instrumental defect or noise at the edge of SH order 14.

\subsubsection{Comparison of 2005 June-July with 2006 December IRS Observations}

The 2005 June-July {\it Spitzer} spectra ({\bf Fig. 5}) were taken more than two months after 
the peak of the 2005 optical outburst. Even so late after the peak, 3C 454.3 was a factor of $\sim 20$ brighter 
at mid-IR wavelengths than at any previously published epoch. The peak flux during this period
reached 1.14 $\pm 0.03$ Jy at 12 $\mu$m and 2.46 $\pm 0.07$ Jy at 24 $\mu$m ({\bf Table} 10). The mean 
spectrum was fit by a power law $F_\nu \sim \nu^{-1.2}$ at 6-12 $\mu$m, with an exponential cutoff 
at longer wavelengths. The primary synchrotron bump peaked at an observed frequency of 1-4$\times10^{13}$
 Hz (30-7 $\mu$m).

In 2006 December, we found 3C 454.3 in a relatively low state ({\bf Fig. 6}). The low state spectrum is 
well fit by a simple power law with spectral index $\alpha=1.3$. We find no evidence of emission lines or 
silicate emission stronger than 3\% of the continuum. This suggests that the low-state IR SED is dominated 
by synchrotron emission from the jet. The flux was 28 $\pm 1$ mJy at 12 $\mu$m, a factor of 40 down from the maximum 
in 2005 July and slightly fainter than the historical ISO measurement of 34 mJy at 12.8 $\mu$m \citep{hmb04}. 

\subsection{Optical Observations}

Since 1988, T. Balonek has conducted blazar monitoring observations with Colgate University's Foggy 
Bottom Observatory. As described by \cite{kb07}, observations are made ``using a Ferson 
16 inch Newtonian/Cassegrain reflecting telescope equipped with a Star I CCD
system. The images range in exposure time from 1 to 5 minutes (most images being
2 minutes) and include the B, V, R, and I filters, designed by \cite{beck89} to conform
to the Johnson-Cousins system...The images were reduced using standard IRAF packages and customized scripts 
written to facilitate the data handling for our system.'' In the case of 3C 454.3, V, R and I band data were 
obtained. The photometry for all of the images was calculated using the IRAF apphot package
with a 16'' diameter aperture and a sky annulus with an inner diameter of
20'' and an outer diameter of 40''.  Some of the Colgate data were previously shown by \cite{Rai08}. 
The Foggy Bottom Observatory data, uncorrected for absorption,  are listed in {\bf Tables 6-8}.
 
Since 2005, the members of the NASA Space Interferometry Mission Key Project on Active Galactic Nuclei, led by 
A. Wehrle, have conducted blazar monitoring observations using the Caltech automated 1.5m telescope at Palomar 
Observatory \citep{cfm06}.  We intensified monitoring of 3C 454.3 when it flared in 2005. Data in filter bands V 
and R were obtained \footnote{Note that the redshift of 3C 454.3 brings the broad emission lines from C III] 
(1909 \AA ~rest, 3549 \AA ~observed) into the blue edge of the Palomar B band filter and Mg II 
(2798 \AA ~rest, 5201 \AA ~observed) into both B and  V Palomar filters.  See \cite{pft05} for Hubble FOS spectra 
of 3C 454.3 taken in 1991 and 1995.}. Data were reduced with IRAF using aperture photometry, with an
4.56" diameter aperture and a sky annulus of 3.8" radius and  3.8" width. The comparison star sequence is given 
online\footnote{http://www.lsw.uni-heidelberg.de/projects/extragalactic/charts/} and originally by \cite{crai77}, 
\cite{an71}, \cite{fi98}, \cite{sb98}, and \cite{Rai98}. 
We used Star H, with V=13.64 and R= 13.10, as the primary comparison star. The Palomar Observatory data,
uncorrected for absorption, are given in {\bf Tables 9 and 10}.

\subsection{Chandra Observations}

Chandra High Energy Transmission Grating (HETG) observations were taken simultaneously with our 
{\it Spitzer} MIPS observations, for 2.145 ks from 21:35-22:42 UT on 2007 January 1. The grating was used to 
reduce potential photon pile up ~effects on the ACIS-S detector. The data were reduced using standard 
point-source grating analysis threads under CIAO 4.0. The MEG and HEG $\pm 1$ order grating spectra were grouped
to have a minimum of 20 counts/bin.  We fixed the hydrogen column density for model fits at 
$N_H=6.5\times 10^{20}$ cm$^{-2}$ \citep{dl90} and assumed standard elemental abundances to estimate
Galactic absorption. The best fit to the X-ray continuum was obtained with a single power law of 
slope $\Gamma = 1.53 \pm 0.12$ and flux at 2-10 keV of $5.8 \pm 0.2\times 10^{-12}$ erg cm$^{-2}$ s$^{-1}$ 
($8.14 \times 10^{-4}$ ph s$^{-1}$ cm$^{-2}$). Adding absorption intrinsic to the host galaxy to the model did 
not improve the fit. Similar fits were obtained using either $\xi^2$ or C-statistics. We report fit parameters
using the latter, because of small photon number statistics in spite of grouping the data.

\section{Results}

\subsection{Optical Light Curves}

The optical light curves are shown in {\bf Figure 9}. We scaled the Colgate V band data  by 1.15 to match 
the Palomar V data in constructing the light curves; the discrepancy is probably due to slight mismatches in filters.
The optical light curves give the appearance of a series of flares superimposed on an exponential fading of the 
big flare in 2005 May, with e-folding time of $\sim$ 60 days.  The source's underlying V band flux faded from 23 mJy 
on about 2005 May 10 to 1.1 mJy on about 2005 September 9.  Superimposed flares, numbering about a dozen between May and 
September, doubled the flux on time scales of 24-48 hours, with similar fade times.

%{\bf Plot sub-mm and optical data on same time scale in Figure 8.}

\subsection{ Infrared Light Curves}

The 2005 June-July infrared light curves for IRS and IRAC are shown together with the optical 
light curves in {\bf Figure 9}. The IRS portions of these curves are shown in more detail in 
{\bf Fig. 10ab}. We compute mean 6-12, 6-24, and 12-24 $\mu$m spectral indices to characterize 
the spectral hardness {\bf Fig. 10c}. The photometric variability is similar in all wavebands, 
with greatest amplitude at the shortest wavelengths. The 6 $\mu$m flux increases by a factor 
of 2 during 2005 July from $240 \pm 20$ mJy at epoch 3, flaring to $470 \pm 30$ mJy at epoch 14. 
All of the dips and peaks in the light curves occur simultaneously at 6, 12, and 24 $\mu$m, to within 
0.5 day. There is no measurable lag.

%A smaller, more gradual rise is seen during 2007 June-July from $210 \pm 20$ mJy at 
%epoch 17 to $310 \pm 20$ mJy at epoch 22.

Significant variations are seen in the spectral index as the flux changes in 2005 June - July 
({\bf Fig. 10}). In particular, as the flux increases from epoch 4-5, the slope becomes softer. The flux 
and spectral slope show only small changes from epoch 5-10. This period is followed by a large flare 
from epochs 12-14, when the flux increases by 85\% at 6 $\mu$m and the spectrum hardens. The flux 
falls and the spectrum softens again in the epoch after the peak of the flare. 

%In 2007 July-August, 
%the spectral index gradually drops from $\alpha=1.05$ to 0.95 as the flux rises over this nine day 
%period. We may be less sensitive to spectral slope changes during this period since the synchrotron 
%peak falls in the middle of the IRS band.

The 6-24 $\mu$m spectral index varies from a maximum of  $1.41 \pm 0.06$ at epoch 9
to a minimum of $1.17 \pm 0.06$ at epoch 13 ({\bf Fig. 10}). The spectral index
uncertainties correspond to the IRS flux repeatability of 2\% at 6 $\mu$m
and 3\% at 24 um, as given in Section 2.2.3. There is an anticorrelation 
between spectral index and 6 $\mu$m flux ({\bf Fig. 11}) during 2005 July. The SED becomes harder 
at high frequencies as the flux increases. The peak of the SED also moves to higher frequency 
as the flux increses: from $1.8-2.4 \times  10^{13}$ Hz to $2.2-4.0\times  10^{13}$ Hz. Whenever the 
6 $\mu$m flux increases (epochs 3-4, 7-8, 9-10, 11-14), the spectral
index decreases. This is a consequence of the larger variability amplitude at 
6 $\mu$m than at 24 $\mu$m. The variability amplitude may be moderated by optical
depth to synchrotron self-absorption, which increases with wavelength.

\subsection{Correlation of Optical and Infrared Light Curves}

An inspection of the optical through infrared light curves ({\bf Fig. 9}) reveals correlated variability
from the V-band to 24 $\mu$m. In particular the epoch 14 flare seen in the IRS light curves 
({\bf Fig. 10}), is seen in all of the optical bands as well. However, a close comparision of
the IRAC/IRS 8.0 $\mu$m band with the R-band light curve ({\bf Fig. 12}) shows significant differences
in flaring activity \footnote{We focus here on the joint IRAC/IRS 8.0 $\mu$m band because it has the largest span 
(26 days) of simultaneous mid-IR coverage, and the R band which has the best optical coverage.}.  
In particular, the 8.0 $\mu$m band shows a bright flare at days 15-20, following the epoch 14 flare, which is much 
less pronounced in the R-band. There are clear differences in the relative amplitudes of several other
smaller flares, which change continuously with wavelength between the IR and optical bands.
This behavior may indicate multiple flaring components in the 3C 454.3 SED which have peak amplitudes
at different wavelengths. (See Section 4.4.)

A cross-correlation of the IRAC/IRS 8.0 $\mu$m and R-band light curves shows a pronounced peak
with correlation amplitude 0.7 at a lag of $0 \pm 0.5$ days ({\bf Fig. 12c}). This confirms the
overall impression that the flaring activity is correlated across the optical and mid-IR bands.
However, the cross-correlation curve is broadened, with a secondary peak at a lag of 4 days.
This lag is similar to the time between the two brightest 8.0 $\mu$m flares, which may produce
an accidental (rather than causal) correlation signal.

\subsection{Mid-Infrared- Optical SEDs}

The mid-infrared to optical spectral energy distributions are shown in {\bf Figure 13}.  From the Figure, we observe 
the following.

1) During the period 30 June -13 July 2005, there was less variation at IRS bands than in VRI bands, but during the 
following ten days, 14 July 2005-25 July, there was more variation at IRAC bands than at VRI, including the 5.8 and 
8.0 $\mu$m bands that overlap in wavelength with the IRS. 

2) The SED slope over mid-infrared through near-infrared to optical bands was negative in both bright (2005) and 
faint (2007) states.  When the source was a factor of 6-10 weaker (in 2006 July and 2006 December) than it was during 
our July 2005 observations, \cite{rvl07} observed that the underlying near-infrared-optical (I-R-V bands) SED 
slope was positive, unlike our 2005 observations. In their data,  the optical emission peaked at V, then decreased in 
Swift B and U bands and rose again at Swift W1 and M2 bands.  These data, taken together, are consistent with the 
Raiteri et al. detection of an underlying "small blue bump" (probably a blend of iron lines, MgII lines and Balmer 
continuum from the broad line region) whose V-band-peaked emission is visible only when the synchrotron emission is 
comparatively faint, and whose signature is overwhelmed by synchrotron emission when the source is flaring 6-10 times
brighter.  Our observations do not cover the Swift bands which show the rise in thermal emission from the accretion disk.  

3) The SED does not rise or fall by a constant amount across the mid-infrared to optical bands from day to day.  
Moreover, the individual band-to-band spectral indices are not the same from day to day.  There is no obvious wavelength 
range over which we see the beginning and ending points of added emission.   That implies the newly injected population 
of  leptons giving rise to the synchrotron emission is broad in energy, but we have to keep in mind that the particles 
radiating at a given frequency are from both old and new populations. Previously injected populations contribute at 
lower frequencies long after that population's high energy electrons no longer radiate at high frequencies, assuming 
no change in the ambient magnetic field.

4) Energy gained in a day at optical bands can be lost in a day, but it takes longer to both gain and lose energy at 
mid-infrared bands.  We see no evidence of asymmetric rise or fall in the mid infrared or optical flares. 

\subsection{Comparison of Outburst and Quiescent X-ray Fluxes and Spectral Indices}

On 19 May 2005, during outburst and a month earlier than our first Spitzer observations, Chandra observed 3C 454.3 for 112 ksec using 
the HRC-LETG, as a Target of Opportunity  within a program on flaring blazars (F. Nicastro, PI; Villata et al. 2006). 
The 0.2-8 keV spectrum was fitted with a power law of photon index $\Gamma = 1.477 \pm 0.017$, with 
$N_H = (1.34 \pm 0.05) \times 10^{21}$ cm$^{-2}$, more than twice the Galactic value (a misprint in Villata et al. 2006 
has been corrected).  During our 1 January 2007 Chandra observations, obtained using the HETG mode when the source was 
quiescent, the 2-10 keV spectral index was quite similar ($\Gamma =1.53 \pm 0.12$) using the Galactic $N_H$ value of 
$6.5 \times 10^{20}$ cm$^{-2}$.  The May 2005 de-absorbed fluxes were $F=(5.5 \pm 0.2) \times 10^{-11}$ erg cm$^{-2}$ 
s$^{-1}$ and $(8.4 \pm 0.2) \times 10^{-11}$ erg cm$^{-2}$ s$^{-1}$  in the 0.2-2 keV and 2-8 keV bands (respectively),
compared to the January 2007 flux at 2-10 keV of $5.8 \pm 0.2\times 10^{-12}$ erg s$^{-1}$.  Hence, the 2-10 keV X-ray flux 
dropped by a factor of $\sim$ 10 while the spectral index remained the same, but the bright state required twice as much 
$N_H$ along the line of sight as the quiescent state.  An {\it Integral} observation (Pian et al. 2006) made 15-18 May 2005 was 
fitted with a power law of index ${\Gamma}= 1.8 \pm 0.1$ using the same \citep{dl90} Galactic $N_H$ value  The flux in the band 3-200 
keV was $5.45 \times 10^{-10}$   erg cm$^{-2}$ s$^{-1}$.

%showing that the spectral index had started to flatten into a presumed gamma-ray peak.

\section{Discussion}

\subsection{The Spectral Energy Distributions: Two Synchrotron Peaks}

We use our multiwavelength photometry and the IRS spectra together in constructing radio-X-ray SEDs ({\bf Figure} 14); 
representative data are listed for convenience in {\bf Tables 11 and 12}. In constructing the SEDs, the photometry was 
corrected for Galactic absorption at V, R and I bands by 0.355, 0.286, and 0.208 magnitudes, respectively.  The Colgate 
V band data were also scaled by 1.15 to match the Palomar V data, as mentioned above. 

During 2005 June 30-July 12 (high state), the primary SED peak fell in the IRS 5-35 $\mu$m  band  ({\bf Fig. 14}). 
Strong variability of the continuum level, slope, and peak frequency are characteristic of synchroton
emission from a relativistic radio jet. The largest variability amplitudes are seen in the IRAC 3.6 $\mu$m and near-IR 
(I) bands and the smallest amplitude at 35 $\mu$m. The turnover and lower variability amplitude at lower frequencies 
may indicate the onset of optically thick synchrotron emission. Without long-wavelength IR coverage, we lost track of 
the primary synchrotron peak location during the July 14-26 IRAC campaign. When the MIPS observations picked up on 
July 27, the IR peak had faded somewhat and shifted to the MIPS 70 $\mu$m band.
 
In the 2006 December 25-2007 January 2 low state, the IR SED peaked in the MIPS 70 $\mu$m FIR band. 
The low-state SED was similar to that reported by \cite{hmb04}, and to other low-state measurements 
from NED published over the period 1979-1995 ({\bf Fig. 14}). One striking difference was the higher 
60 $\mu$m flux measured by \cite{hmb04}, which was $\sim4$ times greater than our MIPS 70 $\mu$m flux, 
but less than the 2005 peak. This is additional evidence that the 60 $\mu$m flux is highly variable, and 
indicates that there was little if any contribution from thermal dust emission during the \cite{hmb04}
observation. It is particularly evident during our low-state {\it Spitzer} observations  ({\bf Fig. 14}) that there are 
two peaks in the radio-IR SED, as seen earlier in the  low-state observations, made by combining ISO data from 1996 Dec 12 
and 1997 Dec 18, of \cite{hmb04}. As we recall from Section 3.2.2, the veracity of the Spitzer 160 micron flux 
measurement in 2007 January is backed up by the detection of a similarly low level in 1997 by ISO. 

During both the low and high states, the primary SED peak occurs at MIR-FIR wavelengths, depending on source brightness. The 
secondary peak occurs in the sub-mm, near 850 $\mu$m. Strong variability of both bumps indicates that they are both composed of 
synchrotron emission from the relativistic jet. At the time of the 2005 July 27 MIPS observations, two bumps are also apparent, 
with the minimum falling somewhere between 160 $\mu$m and 850 $\mu$m. The sharpest contrast between the two peaks is seen during 
the ISO epochs, when the infrared peak was defined by the 0.706 Jy peak at 60 $\mu$m. 

In contrast, non-blazar quasars have two thermal peaks in their SEDs: an ultraviolet peak from the accretion 
disk and an infrared peak due to reprocessing of ultraviolet light by a dusty torus. We have plotted the mean 
\cite{rich2006}  QSO SED, compiled from SDSS, Spitzer, near-IR, GALEX, VLA, and ROSAT data in {\bf Figure 14}, 
redshifted and scaled for comparison to the 3C 454.3 low state. The UV thermal peak is matched to the big and 
small blue bumps observed by XMM OM \citep{rvl07}. The IR bump in the low-state 3C 454.3 SED is significantly 
redder and peaks at a much lower frequency than the IR bump in the mean QSO SED. A thermal SED that peaks at 
70 $\mu$m is difficult to model with a standard dusty torus \citep[e.g.,][]{lsh07}, and would instead require 
that the QSO be deeply buried in an optically thick sphere of cold dust, contrary to the lack of any
silicate absorption in the MIR spectrum and presence of an unobscured Big Blue Bump (BBB) component in the UV SED. 

We can also confidently rule out a significant thermal contribution from cold dust heated by star 
formation in the host galaxy of 3C 454.3. In {\bf Figure 14}, we overplot the SED template constructed for a 
star formation dominated hyper-ULIRG \citep{raw09} with bolometric luminosity $1\times 10^{13} L_\odot$, scaled up 
by a factor of $2.2$ to match the low-state 70 $\mu$m (isotropic) luminosity of $2.9 \times 10^{46}$ erg s$^{-1}$. 
Hyper-ULIRG galaxies are extremely rare in the universe and it would be very unusual for a such a high-luminosity,
radio loud, type 1 quasar to have a starburst dominated FIR SED. Furthermore, the low state 5-70 $\mu$m power law 
with spectral index $-1.3$ ($\S$ 3.2.5) is is much flatter than the spectral index of the template hyper-ULIRG SED. 
We therefore conclude that the jet provides the dominant contribution to the low-state SED at 70 $\mu$m.

\subsection{Interpretation of the 2 Synchrotron Peaks}

The presence of two synchrotron peaks has also been seen or suggested in at least 3 BL Lac objects,
including  BL Lac \citep{rtg03,Rai2009_507}, Mrk 501 \citep{vr99}, and Mrk 421 \citep{donna09b}. 
The frequencies of the two peaks are in the near-IR to optical and far-UV to X-ray bands for BL Lacs,
in contrast to the sub-mm and far-IR for the FSRQ 3C 454.3. \cite{rtg03} suggest four possible explanations
for the double peaks in BL Lac:  1) higher than normal ratio of optical to X-ray extinction, 2) bulk Comptonization, 
3) Klein-Nishina effect on Compton cooling, and 4) two synchrotron emission regions. We can immediately rule 
out the first two explanations for the two peaks in 3C 454.3 because they occur at IR and lower frequencies, 
where extinction is not sufficient and there is no lower energy photon field to bulk-Comptonize to IR frequencies.
While the Klein-Nishina effect can flatten the slope at the high-energy end of the relativistic electron energy 
distribution \citep[e.g.,][]{bmr97}, it can not produce the two distinct synchrotron peaks that we observe in 3C 454.3
(Fig. 14).  We therefore focus on the scenario of two synchrotron emission regions. 

Our observations of 3C 454.3 provide some clues to the relationship between the two (mid-IR  and sub-mm) synchrotron
emission peaks. First, that both peaks occur in both the high and low states, over a period of 6 months
(and also in the 1996-1997 ISO observations) indicates that the jet structures that produce them must persist 
in both the flaring and quiescent jet states.  They are not a transient phenomenon particular to the 2005 jet 
flaring episode. However, further monitoring at shorter timescales is crucial to elucidate the relationship between 
the two peaks. Second, the similar flux intensity (in $\nu F_\nu$) and factor $\sim 40$ increase in the flux of both 
synchrotron peaks indicates that they come from regions or components in the near-side jet that have similar Doppler 
beaming factors ($\delta \sim 10$ from VLBI radio observations; see $\S 1$). 

The very dissimilar peak frequencies ($2\times 10^{11}$ vs. $8\times 10^{13}$ Hz), along with the lack of correlated 
variability at short timescales between the two peaks (Fig. 12) indicates that they are produced in distinct regions of the jet 
with different physical parameters.  The primary mid-IR peak likely comes from a more compact region closer to the base 
of the jet than the sub-mm peak because of synchtrotron self-absorption.  In the 2005 outburst, both regions appear to 
have responded to a major disturbance traveling along the jet. If the two regions are at different radii along the jet, 
then relativistic light travel time effects must shorten the apparent time between the disturbance at the first and 
second regions. The peak of the 2005 optical outburst occured sometime before 9 May 2005 (the start of a 
WEBT observing campaign). The 2005 July sub-mm flare occured sometime between 16 May and 3 June 2005 $ \Delta t \sim 7-25$ days 
($>0.5-2 \times 10^6$ s) after the peak of the optical outburst. Assuming the disturbance travelled down the jet at relativistic 
speed $v$, arriving first at the IR emission region and second at the sub-mm emission region, then we estimate the distance 
between the two regions to be

$ \Delta r = \frac{\beta c \Delta t /(1+z)}{1-\beta \cos \theta}$

\noindent where $\beta = v/c$ and $\theta$ is the jet inclination to the line of 
sight. In the case where the jet lies directly along the line of sight, and $\beta -1 << 1$, this simplifies to
$ \Delta r \simeq 2 \Gamma_{j}^2 c \Delta t / (1+z) $, where the jet bulk Lorentz factor is $\Gamma_{j} = (1-\beta^2)^{-1/2}$.  
Assuming a possible range of $\Gamma_{j}=5-20$ (typical for superluminal blazars) and $\theta \sim 1/\Gamma_{j} = 3-11 \arcdeg$ for 
the jet angle to the line of sight inside the beaming cone, then $\beta=0.980-0.9987$, and $\Delta r = 0.07-5$ pc.

%The apparent superluminal motion of 
%$16 h^{-1} c$ observed in VLBI jet components indicates 

We consider two possible geometries for the two synchrotron emission regions (Fig. 15). The geometry suggested
by \cite{Rai2009_507} for BL Lac is a helical jet, where the emission comes from two blobs moving at different angles 
to the line of sight, corresponding to different Doppler parameters. However, it is not clear in this picture why
there would be two distinct synchrotron blobs rather than 3 or more blobs or a continuous distribution of emission 
along the helix. Rather, the existence of exactly two synchrotron peaks which are present in both the low and high
states suggests a more permanent jet structure.

Another possible geometry for the two synchrotron emission peaks is that the IR peak comes from the base 
of the jet, where it becomes optically thin to mid-IR photons, while the sub-mm peak comes from a recollimation
shock located further down the jet. Such a recollimation shock has been predicted with jet hydrodynamic simulations
\citep{lpm89}, where jet recollimation is caused by magnetic hoop stresses within the jet. This scenario has recently 
been invoked to explain the stationary (yet optically variable) HST 1 synchrotron emission knot observed in
the jet of M87 (Nakamura, Garofolo, \& Meier 2010, in preparation). Alternatively, the sub-mm synchrotron emission
region may mark a location where the jet runs into an external density enhancement such as the broad-line region.

If two distinct synchrotron-emitting populations of particles are moving with 
different speeds and orientations to our line of sight, both populations would have to be folded into models that compute 
the contributions to the spectral energy distribution from synchrotron self-Compton emission, and external Compton emission
from the accretion disk, broad line region, and corona, as has been done for a single broken-power-law-distribution 
population by \cite{ver09}.  The resulting SEDs may show two peaks in the $\gamma$-ray band, corresponding to the two 
peaks in the infrared and sub-mm bands.

\subsection{The Origin of mid-IR Flaring}

Our observations may be further interpreted in the framework of the relativistic shock model 
\citep{mg85}. In this model, shocks are introduced at the base of a pre-existing
relativistic jet by events in the accretion disk or supermassive black hole ergosphere. 
Shocks may be produced by an increase of the relativistic electron energy density, bulk flow 
speed, or changes in the magnetic field configuration. The electron density, energy density, 
and magnetic field will be enhanced at a shock front, leading to copious synchrotron and
inverse Compton emission observed as a bright flare at radio-gamma frequencies.

According to the shock model, the peak of the mid-IR emission comes from a location 
along the relativistic jet where the shock becomes optically thin to synchrotron 
self-absorption. As the shock region progresses along the jet and expands, it will 
become optically thin at progressively lower frequency, and the synchrotron peak 
will move from the mid-IR into the far IR as it fades. The historical SED ({\bf Figure} 14)
shows that a synchrotron peak at $\sim 10^{12}$ Hz ($\sim 300$ $\mu$m, rest) is more 
typical of 3C 454.3 in a non-flaring state.

As the shock propagates down the jet, it loses energy primarily through inverse Compton
scattering. An increase in the ambient photon density field, e.g. flares from the 
accretion disk or close-encounters with broad-line clouds would drain energy 
from the shock via Compton drag. This can not explain mid-IR flaring, though 
it could certainly cause flaring of Compton X-rays and $\gamma$-rays, as for 3C279  
\citep{w98}.

The shock may be re-energized as it runs into inhomogeneities in the jet or 
surrounding interstellar medium and bulk kinetic energy is converted to electron
kinetic energy and magnetic fields. This could explain why we see large-amplitude 
mid-IR flaring with {\it Spitzer} two months after the peak of the 2005 May optical
flare, after the original shock had propagated away from the energy source at the base of 
the jet. 

Variability on longer time scales could also be caused by changes in the Doppler factor as the 
shock velocity changes its angle to the line of sight. As pointed out for BL Lac by \cite{Rai2009_507} 
and for 3C 454.3 by \cite{vra07}, beaming in a curved jet varies along the jet and boosts 
the emission from various parts of the jet which are emitting at different frequencies, 
possibly causing changes in the observed flux and spectral index.

\subsection{Shock Parameters for the IR-Optical Emission Region}

We use simple arguments to derive rough estimates of the physical parameters of the IR-optical
flaring synchrotron emission regions in 2005 July. The source doubled its MIR (6 $\mu$m) flux in 5 days from 
2005 July 8-13 (2.7 days in the host galaxy rest frame), giving a variability brightness temperature of 
$T_{\mathrm var}=3.0\times10^{10}$ K, close to the canonical equipartition value of $5\times10^{10}$ K.
The corresponding variability size is $<3.7\times10^{12}$ cm if the shock has a bulk Doppler parameter of $\delta\sim10$. 
This is smaller than the gravitational radius of the supermassive black hole, $R_G=1.5\times 10^{13} M_8$ cm.

Optical synchrotron emission may be produced when the magnetic field and Lorentz factor of the emitting electrons 
are in the range $B\sim$1-10 Gauss and $\gamma \sim10^{3} - 10^{4}$.  We calculate the e-folding time for synchrotron 
cooling ($t_{cooling}= 2.4 \times 10^{9} \gamma_{4}^{-1} B_{\mu G}^{-2}$ yr, where $\gamma_{4} = \gamma /10^{4} $ 
and  B is in units of microGauss) using Eq. 9.23 of \cite{kro99} for B=10 Gauss and $\gamma =10^{4}$ and $10^{3}$ 
of $\sim$13 minutes and $\sim$ 2 hours, respectively.  Both of these values are consistent with variations on 
timescales of days and an upper limit of one day for the time delay between optical and infrared bands.  The overall 
decline in optical flux on e-folding timescales of six months may be due to adiabatic expansion of a newly emitted 
or newly coalesced jet component.

\section{Conclusions}

We have observed 3C 454.3 with {\it Spitzer} IRS, IRAC, and MIPS, in a high state in 2005 June-July,
two months after the 2005 May optical outburst, and with {\it Spitzer} and {\it Chandra} HETGS in a low state in 
2006 December-2007 January.

We find that:

1) There are no significant narrow or broad emission or absorption features in either 
   the low or high-resolution mid-IR spectra of the high state or low resolution spectrum
   of the low state, consistent with a dominant nonthermal (synchrotron) emission mechanism
   during both states.

2) The mid-IR continuum emission is $\sim$ 30 times brighter than during previous ISO
   observations. The mid-IR flux is highly variable, decreasing by a factor of
   of 20-40 across the 5-35 $\mu$m band between July 2005 and December 2006, indicating Doppler boosted
   synchrotron emission from the jet.

3) There are {\it two} variable synchrotron emission peaks, one in the sub-mm,
   and one in the mid-IR. The lack of correlated variability between the peaks on 
   1-30 day time scales and a lag of 7-25 days between the sub-mm and optical outbursts
   indicates two separate synchrotron emission regions, separated by $\sim 0.07-5$ pc.

4) The frequency of the primary synchrotron peak varies from $<10^{13}$ Hz to 4$\times 10^{13}$ Hz,
   moving through the {\it Spitzer} IRS bandpass. The peak frequency is higher
   in outburst than during the low state, consistent with injection of high-energy
   electrons in a jet shock.

5) The 6-24 $\mu$m  spectral slope becomes harder during flares 
   and the peak of the SED moves to higher frequency. This may indicate that the
   shock is being re-energized as a new jet component runs into inhomogeneities
   in a slower, preexisting jet or surrounding ISM.

6) No time delay between optical and infrared light curves was observed; any time delay is smaller than 0.5 day.
   However, there are significant differences between the flaring behavior in the
   optical and infrared bands. Some flares have larger amplitude in the mid-IR, while others
   have larger amplitude in the optical.

7) The near-IR -optical spectral slope varies dramatically on the timescale of days, 
   perhaps indicating injection of high energy electrons in jet shocks.

\acknowledgements

This work is based on observations made with the {\it Spitzer} Space Telescope, 
which is operated by the Jet Propulsion Laboratory, California Institute of 
Technology under NASA contract 1407. Optical observations at Palomar Observatory were made possible through 
the NASA Space Interferometry Mission Preparatory Science Program for the Key Project on Active Galactic Nuclei. 
Support for this work was provided by NASA through awards issued by JPL/Caltech. The Submillimeter Array is a 
joint project between the Smithsonian Astrophysical Observatory and the Academia Sinica Institute of Astronomy 
and Astrophysics and is funded by the Smithsonian Institution and the Academia Sinica.  This work is partly based  
on data taken and assembled by the WEBT collaboration and stored in the WEBT archive t the Osservatorio Astronomico 
di Torino- INAF (http://www.oato.inaf.it/blazars/webt/). We are grateful to Karl Gordon (Arizona) and Jason Surace 
(Spitzer Science Center) for assistance in processing MIPS and IRAC data, and to Booth Hartley for customizing IPAC 
Skyview for use on Intel Macs. We are grateful to Massimo Villata and Claudia Raiteri for providing WEBT data and 
for helpful discussions, and to Bill Keel (U. of Alabama) and Marco Chiaberge (STScI)  for working with us on the
Hubble images.

%Table 1
\clearpage
\LongTables
\begin{deluxetable}{cccccccc}
\tablecaption{Submillimeter Array Photometry}
\tablewidth{0pt}
\tablehead{\colhead{Year \tablenotemark{a}} 
             & \colhead{Month}
             & \colhead{Day}
             & \colhead{UT}
             & \colhead{MJD}
             & \colhead{Freq (GHz)}
             & \colhead{Flux (Jy)}
             & \colhead{Error (Jy)}
}
\startdata
2005  &  01  &  13  &  02:57  &  53383.121  &  225.6  &   9.7  &   0.5  \\ 
2005  &  04  &  26  &  16:10  &  53486.672  &  340.8  &  18.0  &   1.1 \\ 
2005  &  04  &  27  &  15:57  &  53487.664  &  340.8  &  22.2  &   1.2 \\ 
2005  &  05  &  02  &  05:36  &  53492.234  &  342.5  &  25.0  &   1.4 \\ 
2005  &  05  &  06  &  05:42  &  53496.238  &  225.5  &  25.9  &   1.3 \\ 
2005  &  05  &  07  &  07:08  &  53497.297  &  340.9  &  23.9  &   1.2 \\ 
2005  &  05  &  08  &  07:16  &  53498.305  &  340.8  &  24.1  &   1.2 \\ 
2005  &  05  &  09  &  05:31  &  53499.230  &  340.9  &  24.0  &   1.3 \\
\enddata
\tablenotetext{a}{UT date.}
\end{deluxetable}
\clearpage

%Table 2
\begin{deluxetable}{cccccccc}
\tablecaption{{\it Spitzer} Infrared IRAC and MIPS Observations of 3C 454.3.}
\tablewidth{0pt}
\tablehead{\colhead{Year \tablenotemark{a}} 
             & \colhead{Month}
             & \colhead{Day}
             & \colhead{Instrument}
             & \colhead{Waveband ($\mu$m)}
             & \colhead{Integration Time\tablenotemark{b}}
}
\startdata
2005 & 07 & 14-26 & IRAC & 3.6, 4.5, 5.8, 8 & $1\times 32$ s \\
2005 & 07 & 27 (twice\tablenotemark{c}) & MIPS & 24, 70, 160  & $(2,2,4) \times 3$ s \\
2006 & 12 & 25 & IRAC & 3.6, 4.5, 5.8, 8 & $1\times 32$ s\\
2007 & 01 & 02  & MIPS & 24, 70  & $2 \times 3$  s\\
\enddata
\tablenotetext{a}{UT date.}
\tablenotetext{b}{Number of exposures or cycles $\times$ exposure time (s).}
\tablenotetext{c}{Observed at 01:48 UT and 15:11 UT}

\end{deluxetable}

%Table 3
\begin{deluxetable}{cccccccc}
\tablecaption{{\it Spitzer} IRS Observations of 3C 454.3.}
\tablewidth{0pt}
\tablehead{\colhead{Epochs}  
             & \colhead{year \tablenotemark{a}} 
             & \colhead{month}
             & \colhead{day}
             & \colhead{SL \tablenotemark{b}} 
             & \colhead{LL \tablenotemark} 
             & \colhead{SH \tablenotemark} 
             & \colhead{LH \tablenotemark}
}
\startdata
1     & 2005 & 06  & 30    & $4\times60$  &  $2\times120$ & $4\times120$ & $4\times60$ \\
2-14  & 2005 & 07  & 01-12 & $4\times60$  &  $2\times120$ & $0$          & $0$ \\
15    & 2005 & 07  & 13    & $4\times60$  &  $2\times120$ & $4\times120$ & $4\times60$ \\
16    & 2006 & 12  & 20    & $16\times14$ & $24\times14$  & $0$          & $0$ \\
%17    & 2007 & 07  & 28    & $16\times14$ & $24\times14$  & $0$          & $0$  \\
%18-22 & 2007 & 08  & 02-06 & $16\times14$ & $24\times14$  & $0$          & $0$  \\

\enddata
\tablenotetext{a}{UT date.}
\tablenotetext{b}{Number of exposures $\times$ exposure time (s).}

\end{deluxetable}

%Table 4
\begin{deluxetable}{ccccccccc}
\tablecaption{{\it Spitzer} Infrared IRAC Photometry of 3C 454.3.}
\tablewidth{0pt}
\tablehead{\colhead{Year \tablenotemark{a}} 
             & \colhead{Month}
             & \colhead{Day}
             & \colhead{UT}
             & \colhead{JD}
             & \colhead{$S_{3.6}[Jy]$\tablenotemark{b}} 
             & \colhead{$S_{4.5}[Jy]$\tablenotemark{c}} 
             & \colhead{$S_{5.8[Jy]}$\tablenotemark{d}} 
             & \colhead{$S_{8.0}[Jy]$\tablenotemark{e}} 
}
\startdata

2005 	&	7	&	14	&	19:41	&	2453566.32014	&	0.172	&	0.247	&	0.37		&	0.616	\\
2005	 	&	7	&	15	&	17:52	&	2453567.24444	&	0.201	&	0.292	&	0.445	&	0.765	\\
2005 	&	7	&	16	&	19:41	&	2453568.32014	&	0.197	&	0.282	&	0.426	&	0.723	\\
2005	 	&	7	&	17	&	19:08	&	2453569.29722	&	0.239	&	0.329	&	0.481	&	0.774	\\
2005 	&	7	&	18	&	21:03	&	2453570.37708	&	0.206	&	0.284	&	0.417	&	0.664	\\
2005		&	7	&	19	&	21:17	&	2453571.38681	&	0.185	&	0.258	&	0.382	&	0.610		\\
2005		&	7	&	20	&	20:08	&	2453572.33889	&	0.163	&	0.228	&	0.335	&	0.533	\\
2005	 	&	7	&	21	&	20:51	&	2453573.36875	&	0.133	&	0.187	&	0.277	&	0.443	\\
2005		&	7	&	22	&	18:49	&	2453574.28403	&	0.102	&	0.145	&	0.216	&	0.347	\\
2005	 	&	7	&	23	&	20:29	&	2453575.35347	&	0.092	&	0.128	&	0.187	&	0.293	\\
2005 	&	7	&	24	&	08:28	&	2453575.85278	&	0.097	&	0.135	&	0.198	&	0.309	\\
2005 	&	7	&	25	&	19:49	&	2453577.32569	&	0.113	&	0.156	&	0.226	&	0.349	\\
2006 	&	12	&	25	&	06:23	&	2454081.76597	&	5.65E-3	&	7.30E-3	&	9.10E-3	&	14.2E-3	\\
\enddata
\tablenotetext{a}{UT date.}
\tablenotetext{b}{Flux at $3.6 \mu$m; error is 5\%.}
\tablenotetext{c}{Flux at $4.5 \mu$m; error is 5\%.}
\tablenotetext{d}{Flux at $5.8 \mu$m; error is 5\%.}
\tablenotetext{e}{Flux at $8.0 \mu$m; error is 5\%.}
\end{deluxetable}
\clearpage

%Table 5
\begin{deluxetable}{cccccccccccc}
\tablecaption{{\it Spitzer} IRS mid-IR spectrophotometry of 3C 454.3.}
\tablewidth{0pt}
\tablehead{\colhead{Epoch}  
             & \colhead{day\tablenotemark{a}} 
             & \colhead{$F_6$\tablenotemark{b}} 
             & \colhead{$+/-$}
             & \colhead{$F_{12}$\tablenotemark{c}}
             & \colhead{$+/-$}
             & \colhead{$F_{18}$\tablenotemark{d}}
             & \colhead{$+/-$}
             & \colhead{$F_{24}$\tablenotemark{e}}
             & \colhead{$+/-$}
             & \colhead{$F_{30}$\tablenotemark{f}}
             & \colhead{$+/-$}
}
\startdata
  1 &   0.0721 & 0.29 & 0.02 & 0.74 & 0.03 & 1.26 & 0.04 & 1.80 & 0.05 & 2.24 & 0.09 \\
  2 &   1.2601 & 0.27 & 0.02 & 0.68 & 0.02 & 1.12 & 0.04 & 1.57 & 0.04 & 1.91 & 0.07 \\
  3 &   2.2021 & 0.24 & 0.02 & 0.62 & 0.02 & 1.03 & 0.03 & 1.45 & 0.04 & 1.79 & 0.07 \\
  4 &   2.9945 & 0.29 & 0.02 & 0.74 & 0.02 & 1.22 & 0.04 & 1.67 & 0.04 & 2.04 & 0.07 \\
  5 &   4.0043 & 0.29 & 0.02 & 0.79 & 0.03 & 1.36 & 0.05 & 1.95 & 0.06 & 2.45 & 0.10 \\
  6 &   5.4844 & 0.27 & 0.02 & 0.72 & 0.03 & 1.27 & 0.04 & 1.86 & 0.06 & 2.34 & 0.09 \\
  7 &   6.3560 & 0.26 & 0.02 & 0.71 & 0.03 & 1.25 & 0.04 & 1.81 & 0.06 & 2.28 & 0.09 \\
  8 &   7.1905 & 0.28 & 0.02 & 0.76 & 0.03 & 1.30 & 0.05 & 1.86 & 0.06 & 2.34 & 0.09 \\
\enddata
\tablenotetext{a}{Start time of exposure (days) from UT 2005 June 30 00:00 (MJD 53551).}
\tablenotetext{b}{Mean 5.5-6.5 $\mu$m (3.0-3.5 $\mu$m rest) flux density (Jy).}
\tablenotetext{c}{Mean 11.5-12.5 $\mu$m (6.2-6.7 $\mu$m rest) flux density (Jy).}
\tablenotetext{d}{Mean 17-19 $\mu$m (9.1-10.2 $\mu$m rest) flux density (Jy).}
\tablenotetext{e}{Mean 23-25 $\mu$m (12.4-13.4 $\mu$m rest) flux density (Jy).}
\tablenotetext{f}{Mean 28-32 $\mu$m (15.1-17.2 $\mu$m rest) flux density (Jy).}
\end{deluxetable}

%Table 6
\begin{deluxetable}{ccccccccc}
\tablecaption{Foggy Bottom Observatory V-band Observations of 3C 454.3.}
\tablewidth{0pt}
\tablehead{\colhead{Year}  
             & \colhead{Month}
             & \colhead{Day}
             & \colhead{UT} 
             &\colhead{Julian Date}
             & \colhead{V mag} 
             & \colhead{V mag error} 
             & \colhead{Flux [mJy]}
             &\colhead{Flux error [mJy]}
}
\startdata
2005	&	5	&	9	&	8:42	&	2453499.863	&	12.33	&	0.031	&	36.01	&	1.112	\\
2005	&	5	&	10	&	8:39	&	2453500.861	&	12.804	&	0.031	&	23.269	&	0.713	\\
2005	&	5	&	13	&	8:31	&	2453503.855	&	13.264	&	0.032	&	15.242	&	0.484	\\
2005	&	5	&	17	&	8:19	&	2453507.847	&	13.585	&	0.035	&	11.337	&	0.397	\\
2005	&	5	&	19	&	7:57	&	2453509.832	&	12.697	&	0.029	&	25.697	&	0.734	\\
2005	&	5	&	21	&	8:02	&	2453511.835	&	13.509	&	0.034	&	12.159	&	0.414	\\
2005	&	5	&	26	&	7:43	&	2453516.822	&	13.657	&	0.027	&	10.606	&	0.29	\\
2005	&	5	&	28	&	8:09	&	2453518.84	&	13.868	&	0.047	&	8.741	&	0.412	\\
\enddata
\end{deluxetable}

%Table 7
\begin{deluxetable}{ccccccccc}
\tablecaption{Foggy Bottom Observatory R-band Observations of 3C 454.3.}
\tablewidth{0pt}
\tablehead{\colhead{Year}  
             & \colhead{Month}
             & \colhead{Day}
             & \colhead{UT} 
             &\colhead{Julian Date}
             & \colhead{R mag} 
             & \colhead{R mag error} 
             & \colhead{Flux [mJy]}
             &\colhead{Flux error [mJy]}
}
\startdata
2004	&	6	&	21	&	7:41	&	2453177.82	&	15.315	&	0.019	&	2.304	&	0.044	\\
2004	&	7	&	7	&	7:25	&	2453193.809	&	15.189	&	0.048	&	2.588	&	0.124	\\
2004	&	7	&	21	&	7:21	&	2453207.806	&	15.268	&	0.052	&	2.406	&	0.126	\\
2004	&	8	&	22	&	4:45	&	2453239.698	&	15.66	&	0.042	&	1.677	&	0.071	\\
2004	&	8	&	25	&	7:00	&	2453242.792	&	15.92	&	0.134	&	1.32	&	0.176	\\
2004	&	9	&	5	&	3:27	&	2453253.644	&	14.803	&	0.029	&	3.692	&	0.108	\\
2004	&	9	&	12	&	7:05	&	2453260.796	&	15.212	&	0.045	&	2.534	&	0.113	\\
2004	&	9	&	14	&	4:40	&	2453262.695	&	14.419	&	0.017	&	5.261	&	0.087	\\
\enddata
\end{deluxetable}

%Table 8
\begin{deluxetable}{ccccccccc}
\tablecaption{Foggy Bottom Observatory I-band Observations of 3C 454.3.}
\tablewidth{0pt}
\tablehead{\colhead{Year}  
             & \colhead{Month}
             & \colhead{Day}
             & \colhead{UT} 
             &\colhead{Julian Date}
             & \colhead{I mag} 
             & \colhead{I mag error} 
             & \colhead{Flux [mJy]}
             &\colhead{Flux error [mJy]}
}
\startdata
2005	&	5	&	9	&	8:49	&	2453499.868	&	11.22	&	0.018	&	100.12	&	1.828	\\
2005	&	5	&	10	&	8:47	&	2453500.867	&	11.541	&	0.019	&	74.467	&	1.412	\\
2005	&	5	&	13	&	8:42	&	2453503.863	&	12.157	&	0.017	&	42.259	&	0.715	\\
2005	&	5	&	17	&	8:30	&	2453507.854	&	12.367	&	0.012	&	34.822	&	0.432	\\
2005	&	5	&	19	&	7:40	&	2453509.82	&	11.522	&	0.009	&	75.806	&	0.664	\\
2005	&	5	&	19	&	8:26	&	2453509.852	&	11.529	&	0.01	&	75.317	&	0.775	\\
2005	&	5	&	21	&	8:17	&	2453511.846	&	12.309	&	0.01	&	36.709	&	0.383	\\
2005	&	5	&	26	&	7:57	&	2453516.832	&	12.445	&	0.01	&	32.399	&	0.322	\\
\enddata
\end{deluxetable}

\clearpage

%Table 9
\begin{deluxetable}{ccccccccc}
\tablecaption{Palomar Observatory V-band Observations of 3C 454.3.}
\tablewidth{0pt}
\tablehead{\colhead{Year}  
             & \colhead{Month}
             & \colhead{Day}
             & \colhead{UT} 
             &\colhead{Julian Date}
             & \colhead{V mag} 
             & \colhead{V mag error} 
             & \colhead{Flux [mJy]}
             &\colhead{Flux error [mJy]}
}
\startdata
2005	&	05	&	12	&	11:40	&	2453502.9861	&	13.271	&	0.002	&	17.894	&	0.033	\\
2005	&	05	&	13	&	11:40	&	2453503.9861	&	13.410	&	0.002	&	15.744	&	0.029	\\
2005	&	05	&	16	&	10:54	&	2453506.9542	&	13.414	&	0.002	&	15.686	&	0.029	\\
2005	&	05	&	18	&	11:37	&	2453508.9840	&	13.074	&	0.002	&	21.454	&	0.040	\\
2005	&	05	&	20	&	11:11	&	2453510.9660	&	13.340	&	0.002	&	16.792	&	0.031	\\
2005	&	05	&	23	&	10:34	&	2453513.9403	&	13.854	&	0.004	&	10.459	&	0.039	\\
2005	&	05	&	24	&	11:10	&	2453514.9653	&	13.521	&	0.003	&	14.214	&	0.039	\\
2005	&	05	&	30	&	10:06	&	2453520.9208	&	14.139	&	0.004	&	8.045	&	0.030	\\
\enddata
\end{deluxetable}

%Table 10
\begin{deluxetable}{ccccccccc}
\tablecaption{Palomar Observatory R-band Observations of 3C 454.3.}
\tablewidth{0pt}
\tablehead{\colhead{Year}  
             & \colhead{Month}
             & \colhead{Day}
             & \colhead{UT} 
             & \colhead{Julian Date}
             & \colhead{R mag} 
             & \colhead{R mag error} 
             & \colhead{Flux [mJy]}
             &\colhead{Flux error [mJy]}
}
\startdata
2005	&	5	&	12	&	11:40	&	2453502.9861	&	12.721	&	0.002	&	25.128	&	0.046	\\
2005	&	5	&	13	&	11:40	&	2453503.9861	&	12.860	&	0.002	&	22.108	&	0.041	\\
2005	&	5	&	16	&	10:54	&	2453506.9542	&	12.864	&	0.002	&	22.027	&	0.041	\\
2005	&	5	&	18	&	11:37	&	2453508.9840	&	12.524	&	0.002	&	30.127	&	0.056	\\
2005	&	5	&	20	&	11:11	&	2453510.9660	&	12.790	&	0.002	&	23.580	&	0.043	\\
2005	&	5	&	23	&	10:34	&	2453513.9403	&	13.304	&	0.004	&	14.688	&	0.054	\\
2005	&	5	&	24	&	11:10	&	2453514.9653	&	12.971	&	0.003	&	19.960	&	0.055	\\
2005	&	5	&	30	&	10:06	&	2453520.9208	&	13.589	&	0.004	&	11.297	&	0.042	\\
\enddata
\end{deluxetable}

%Table 11
\begin{deluxetable}{cccccccc}
\tablecaption{Representative Multiwavelength Spectrum in High State }
\tablewidth{0pt}
\tablehead{\colhead{Year} 
             & \colhead{Month}
             & \colhead{Day}
             & \colhead{MJD}
             & \colhead{Band}
             & \colhead{Freq (Hz)}
             & \colhead{Flux (Jy)}
             & \colhead{Error (Jy)}
}
\startdata

%2005  &  06  &  30    &  53551.355  &  225.6 GHz &   $2.256\times 10^{11}$ & 32.1  &   1.6 \\ 
%2005  &  07  &  01    &  53552.223  &  348.7 GHz &    $3.487\times 10^{11}$ &27.2  &   1.4 \\ 
2005  &  07  &  04  &   53555.555  &  221.5 GHz  &  $2.215\times 10^{11}$  & 41.0  &   2.0 \\ 
2005  &  07  &  05   &  53556.031  &  348.7 GHz &  $3.487\times 10^{11}$  & 40.4  &   2.0 \\ 

2007 & 07 & 27   &  53578.6329211 & 24$\mu m$ & $1.25 \times 10^{13}$ & 1.12 & 0.11 \\
2007 & 07 & 27   &  53578.6329211 & 70$\mu m$ & $4.29  \times 10^{12}$ & 3.51 & 0.35 \\
2007 & 07 & 27   &  53578.6329211 & 160$\mu m$ & $1.88  \times 10^{12}$  &  5.43 & 1.09 \\

2005 & 07	& 17 & 53569.29722 & 3.6$\mu m$ & $8.33  \times 10^{13}$ & 0.239 & 1.2E-2 \\
2005 & 07	& 17 & 53569.29722 & 4.5$\mu m$ & $6.67  \times 10^{13}$  & 0.329 & 1.6E-2 \\
2005 & 07	& 17 & 53569.29722 & 5.8$\mu m$ &$ 5.17 \times 10^{13}$  & 0.481&  2.4E-2 \\
\enddata
\tablenotetext{a}{The V,R and I band photometry was 
corrected for Galactic absorption by factors of 1.39, 1.30, and 1.21, corresponding to magnitudes of  0.355, 0.286, and 0.208 , respectively.}
\end{deluxetable}

%Table 12
\begin{deluxetable}{cccccccc}
\tablecaption{Representative Multiwavelength Spectrum in Low State}
\tablewidth{0pt}
\tablehead{\colhead{Year} 
             & \colhead{Month}
             & \colhead{Day}
             & \colhead{MJD}
             & \colhead{Band}
             & \colhead{Freq (Hz)}
             & \colhead{Flux (Jy)}
             & \colhead{Error (Jy)}
}
\startdata

2006  &  12  &  22 &  54091.309  &  225.5 GHz  &  $2.255\times 10^{11}$ &  4.0  &   0.2 \\ 
2006  &  12  &  29   &  54098.270  &  335.3 GHz  &   $3.353\times 10^{11}$ & 2.8  &   0.2 \\ 

2007 & 01 & 01 & 54101.9257388 & 24$\mu m$ & $1.25 \times 10^{13}$ & 0.060 & 0.006 \\
2007 & 01 & 01 & 54101.9257388 & 70$\mu m$ & $4.29  \times 10^{12}$ & 0.189 & 0.019 \\
2007 & 01 & 01 & 54101.9257388 & 160$\mu m$ & $1.88  \times 10^{12}$  &  0.22 & 0.02 \\

2006 	&	12	&	25	&	54081.76597	&	  3.6$\mu m$ & $8.33  \times 10^{13}$ & 5.65E-3 & 2.8E-4 \\
2006 	&	12	&	25	&	54081.76597	&	   4.5$\mu m$ & $6.67  \times 10^{13}$  & 7.30E-3 & 3.7E-4 \\
2006 	&	12	&	25	&	54081.76597	&	  5.8$\mu m$ &$ 5.17 \times 10^{13}$  & 9.10E-3 & 4.5E-4 \\
\enddata

 \tablenotetext{a} {Optical and ultraviolet data (B,U, UVW1, UVM2) are from \cite{rvl07}, their Table 3. Some of the optical data
                    used in this paper (not listed here), including data from December 2006, are stored in the WEBT archive; for 
                    questions regarding their availability, please contact the WEBT President Massimo Villata ({\tt villata@oato.inaf.it})} 
 \tablenotetext{b}{ {\it Chandra} 2-10 keV flux was $5.8 \times 10^{-12}$ erg cm$^{-2}$ sec$^{-1}$ with $\Gamma = 1.53 \pm 0.12$.}
\end{deluxetable}

\clearpage

%Figure 1
\begin{figure}[ht]
  \plotone{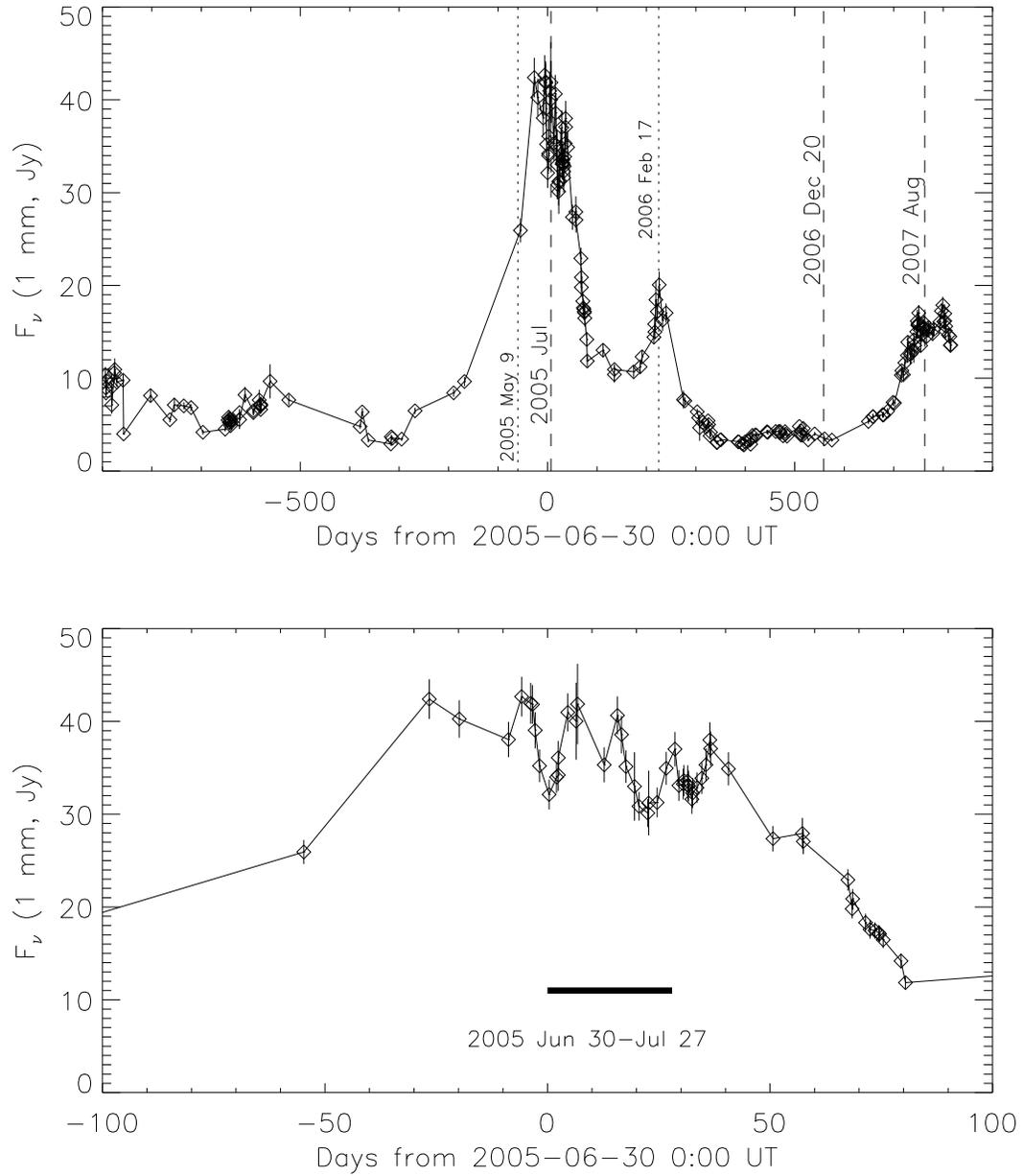}
  \figcaption{Radio (1.3 mm $=$ 230 GHz) variability of 3C 454.3.
              Top: Light curve from 2002-2007. The date of the earliest WEBT observation of the 2005 May optical outburst is 
              indicated by the dotted line. The dates of our Spitzer observations are indicated by dashed lines. Bottom: 
              2005 outburst with the period of the first {\it Spitzer} monitoring campaign indicated by the solid bar.}
\end{figure} 

%Figure 2
\begin{figure}[ht]
 \plotone{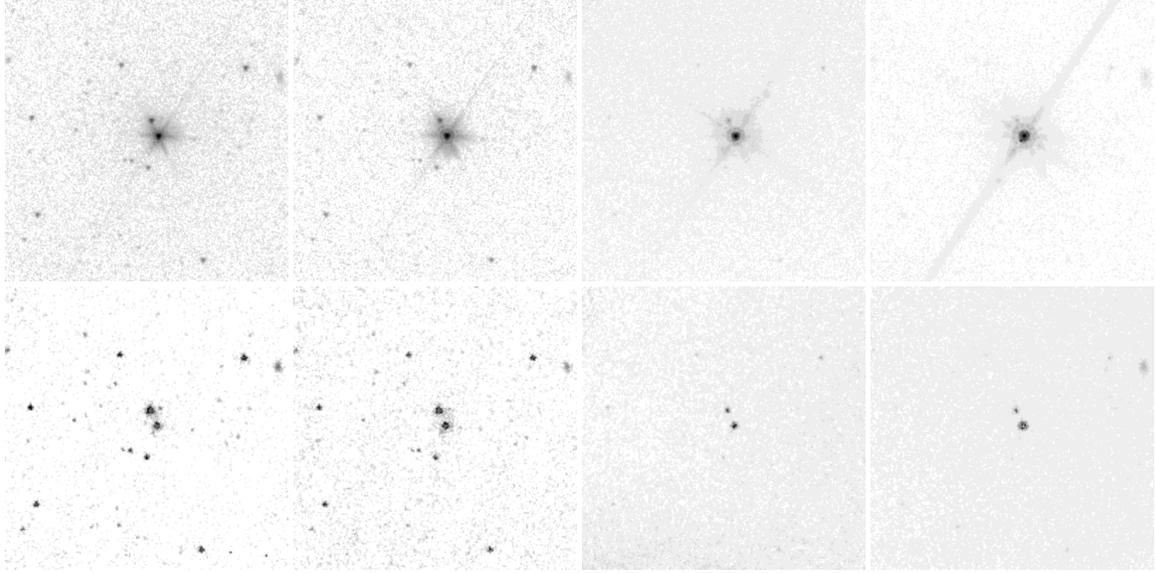}
 \figcaption{{\it Spitzer} IRAC images at 3.6, 4.5, 5.8 and 8 $\mu$m during the bright state on 2005 July 17
             (top four panels) and during the faint state on  2006 December 25 (bottom four panels). Only the central 
             $8\times8$ arcminutes of each image are shown (North up, East left). The transfer
             functions of each panel have been chosen individually to show the most detail.}
\end{figure}

%Figure 3 Hubble and IRAC
\begin{figure}[ht]
 \plotone{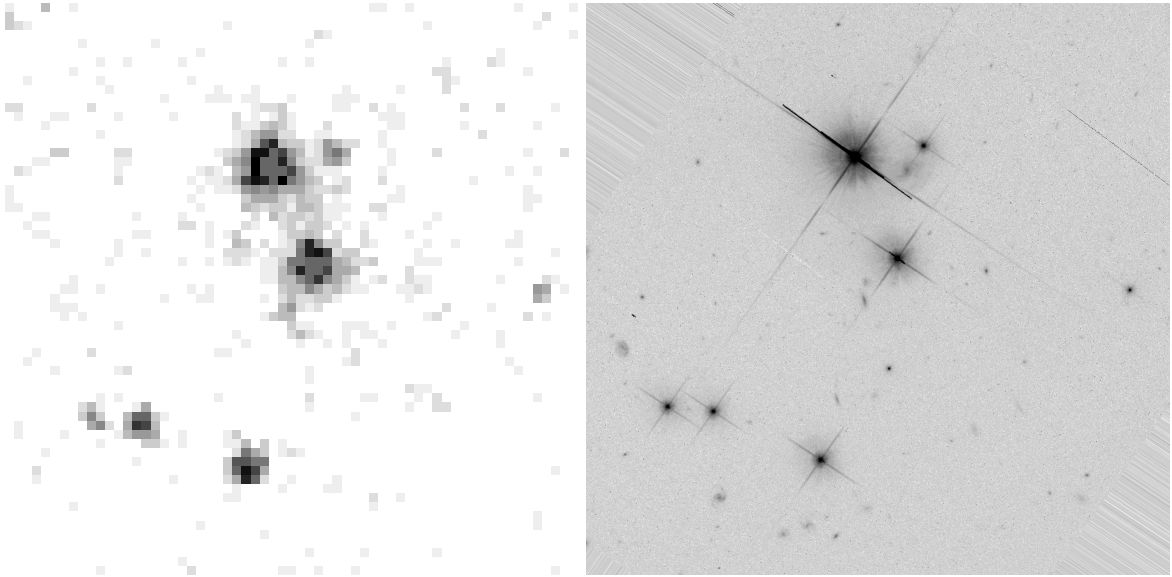}
 \figcaption{Left: {\it Spitzer} IRAC image at 3.6 $\mu$m during the faint state on  2006 December 25. Right: {\it Hubble} 
              image in the F814 filter (2004 August 4, provided by M. Chiaberge). Only the central $1.24\times1.24$ arcminutes 
              of each image are shown (North up, East left). The extension to the SE of 3C 454.3 in the IRAC image
              corresponds to two faint galaxies in the {\it Hubble} image.}
\end{figure}

%Figure 4
\begin{figure}[ht]
  \plotone{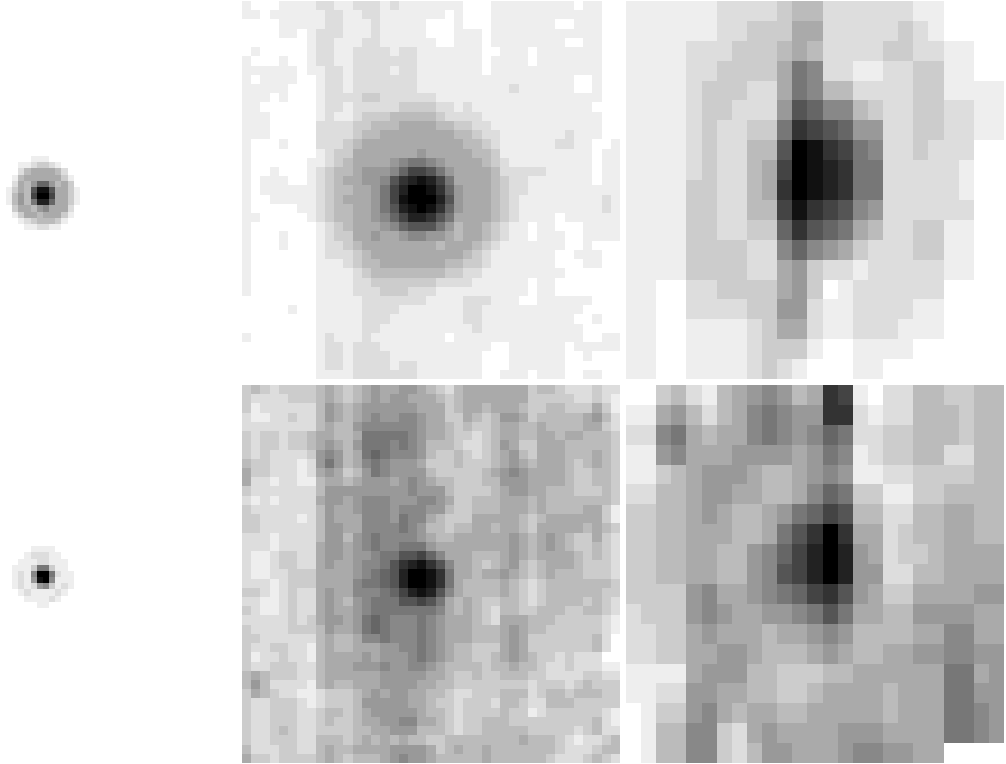}
  \figcaption{{\it Spitzer} MIPS images at 24 $\mu$m (left), 70 $\mu$m (middle), and 160 $\mu$m (right)
              during the bright state on 2005 July 27 (top) and during the faint state on 2007 Jan 1 (bottom).
              Only the centers of each image are shown (in detector coordinates). The transfer functions of each panel 
              have been chosen individually to bring out PSF details. }
\end{figure}
\begin{figure}[ht]
  \plotone{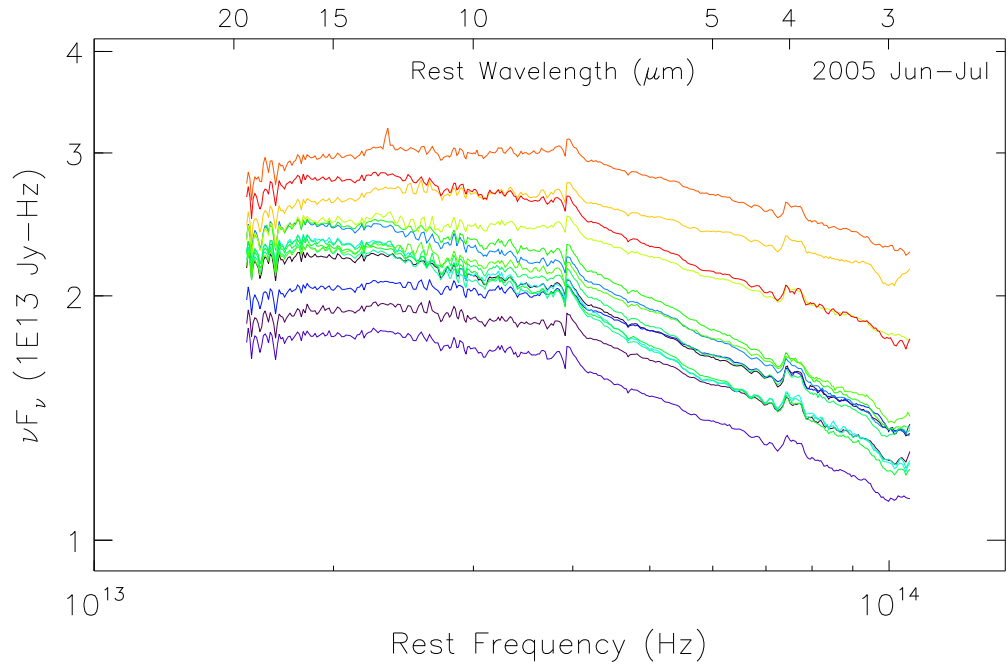}
  \figcaption{{\it Spitzer} IRS low resolution spectra from 2005 June-July. The bumps at  
              $4 \times 10^{13}$ Hz and $7.3 \times 10^{13}$ Hz (14 $\mu$m and 7 $\mu$m, observed) 
              and the dip at  $1 \times 10^{14}$ Hz ($\sim 5.5$ $\mu$m, observed) are instrumental 
              features. The noise spike at $2.2 \times 10^{13}$ Hz (top spectrum) is an artifact of saturation.
              Other small bumps and wiggles in the spectra are attributable to residual detector fringes.}
\end{figure} 

\begin{figure}[ht]
  \plotone{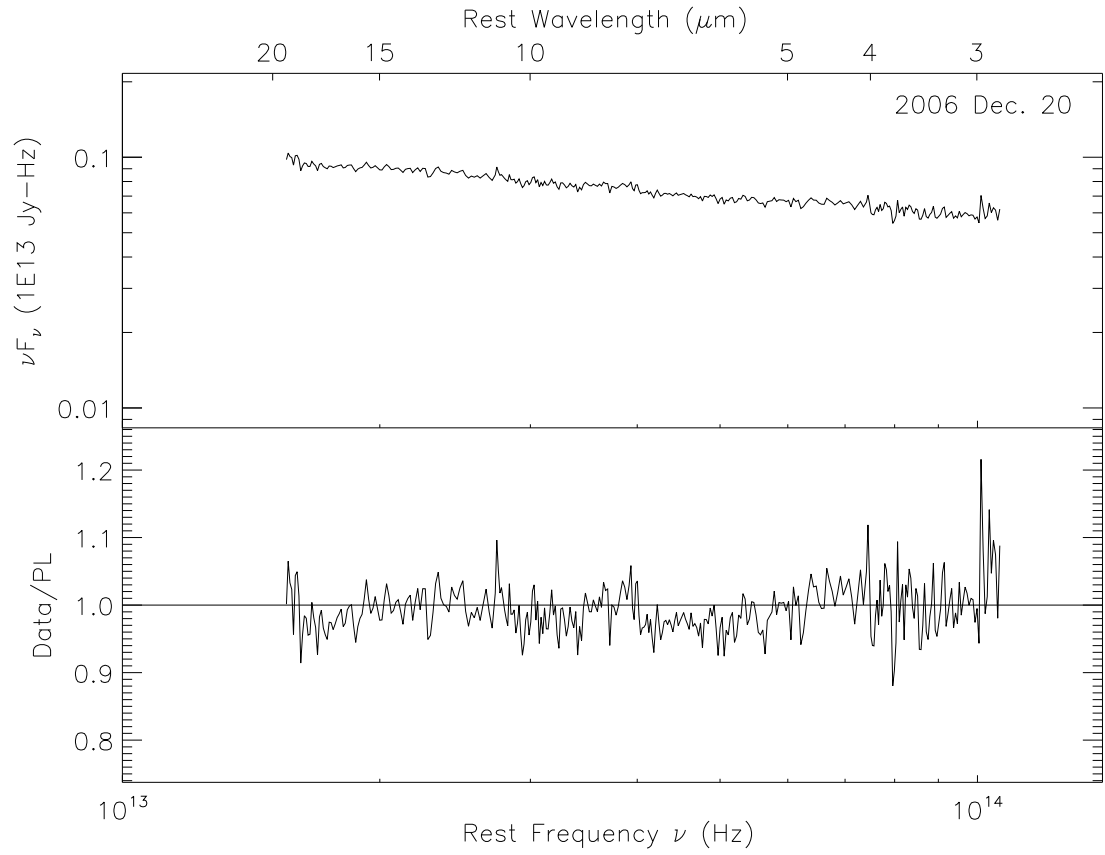}
  \figcaption{Top: {\it Spitzer} IRS low resolution spectrum from 2006 December 20 low state.
              Bottom: Spectrum divided by best fit powerlaw ($\alpha=1.28$).}
\end{figure} 

\begin{figure}[ht]
  \plotone{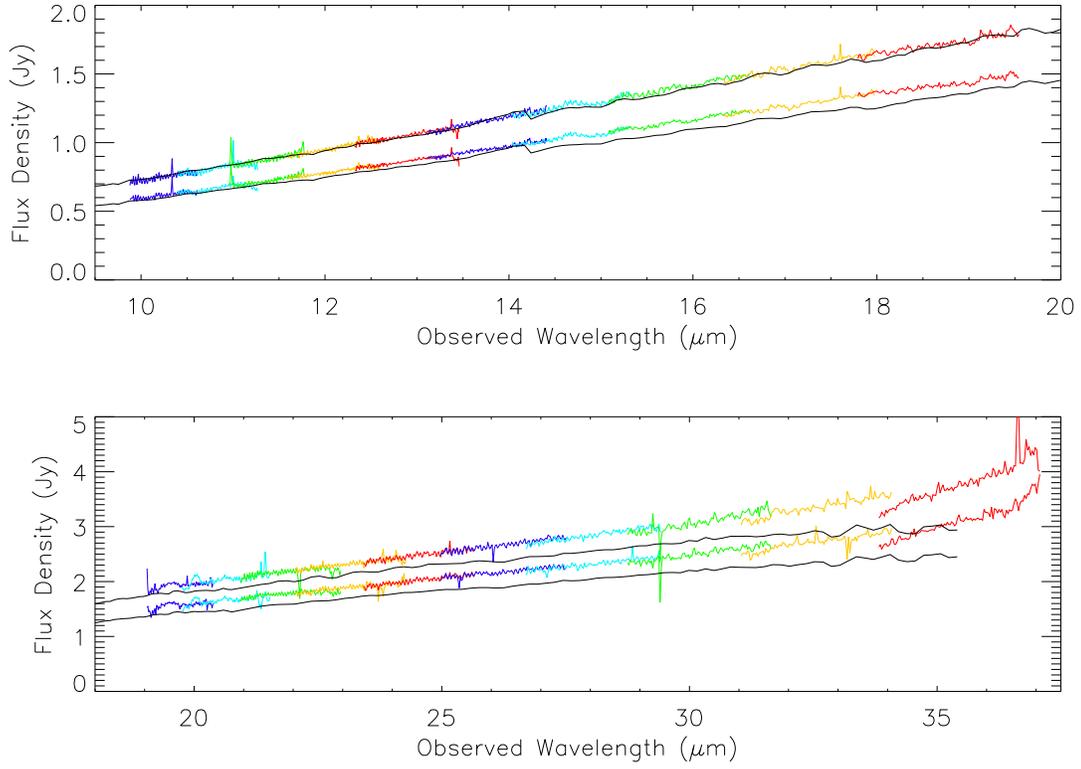}
  \figcaption{High resolution spectra of 3C 454.3 (2005 epochs 1 and 15), with low resolution spectra
              (black lines) from the same epochs overlaid. Background has been subtracted for the low resolution spectra but not for 
              the high resolution spectra (for which we had no separate background measurements), hence, there is an offset between 
              the two spectra. Top panel: SH; bottom panel: LH.}
\end{figure}

\begin{figure}[ht]
  \plotone{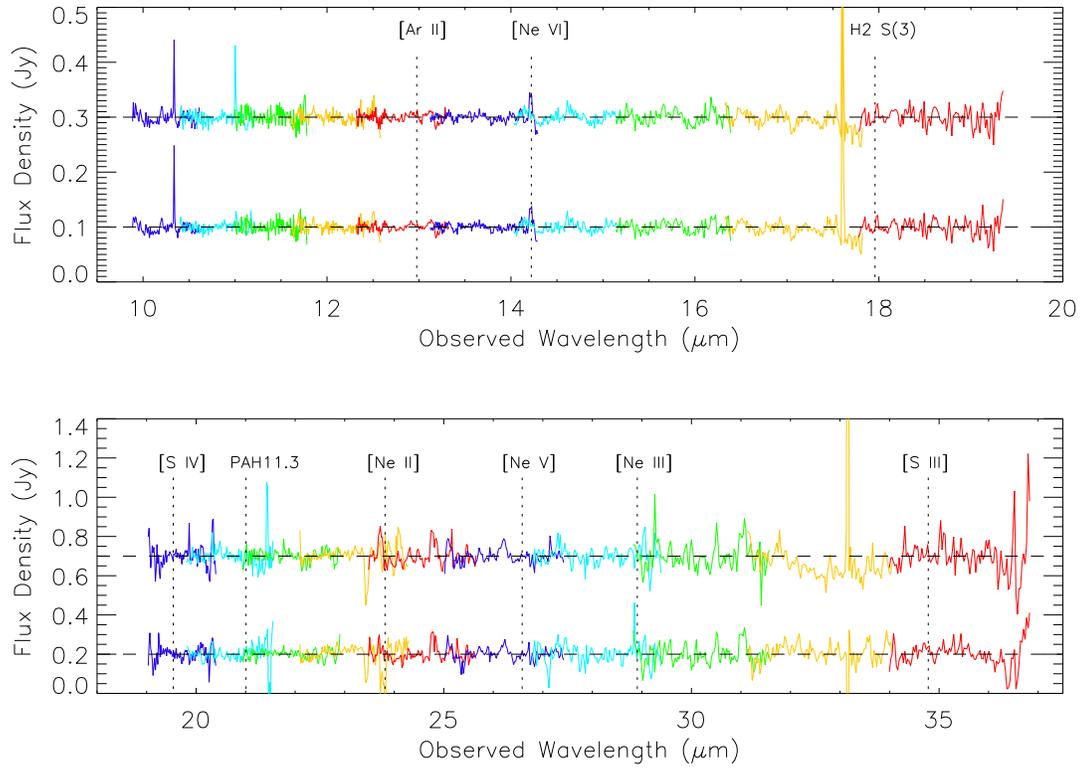}
  \figcaption{Continuum-subtracted high resolution spectra of 3C 454.3. The 2005 epochs (1 and 15)
              are both offset from zero by an arbitrary amount for clarity. Top panel: SH; bottom panel: LH.}
\end{figure} 

\begin{figure}[ht]
  \plotone{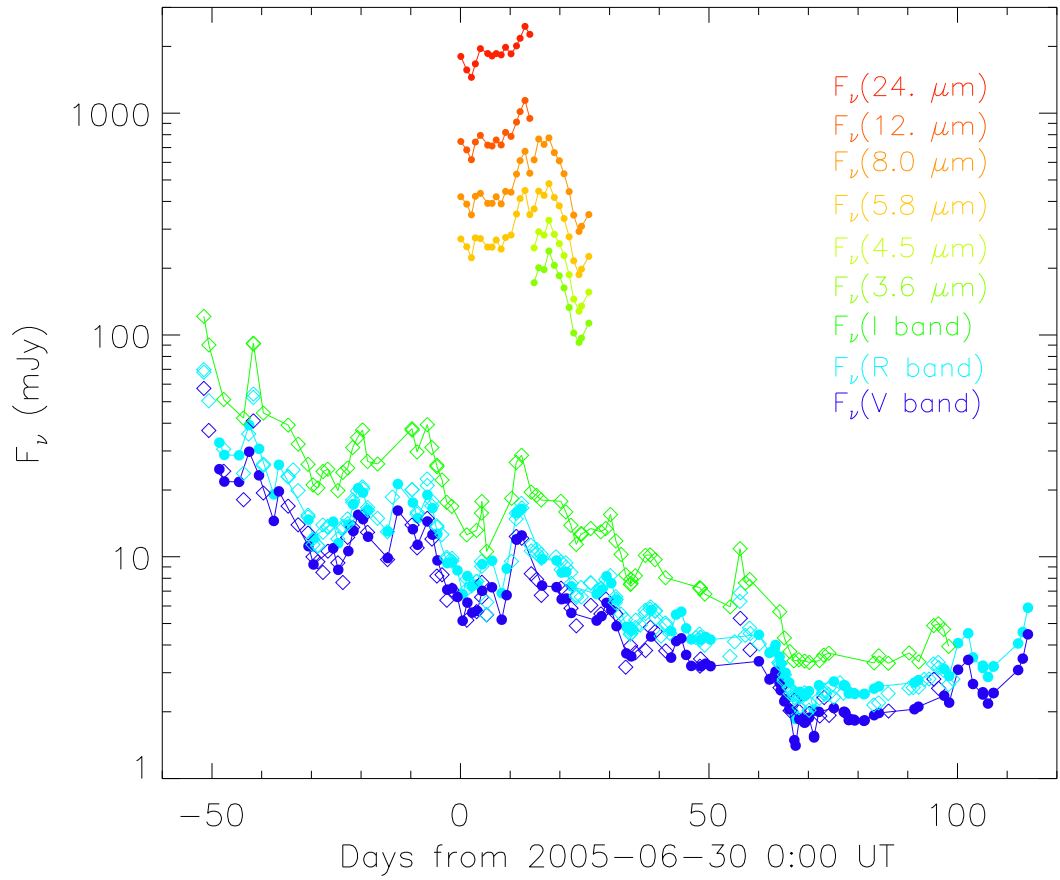}
  \figcaption{Optical and mid-IR light curves for 2005 May-October. Optical V, R, and I band data from 
              Foggy Bottom Observatory and Palomar Observatory are corrected for Galactic extinction and 
              reddening.  Mid-IR data are from the 2005 June 30-July 26 {\it Spitzer} IRS and IRAC campaigns. 
              The optical fluxes showed flaring behavior superposed on an overall decline. Measurement uncertainties are generally smaller than the plot symbols.}
\end{figure} 

\begin{figure}[ht]
  \plotone{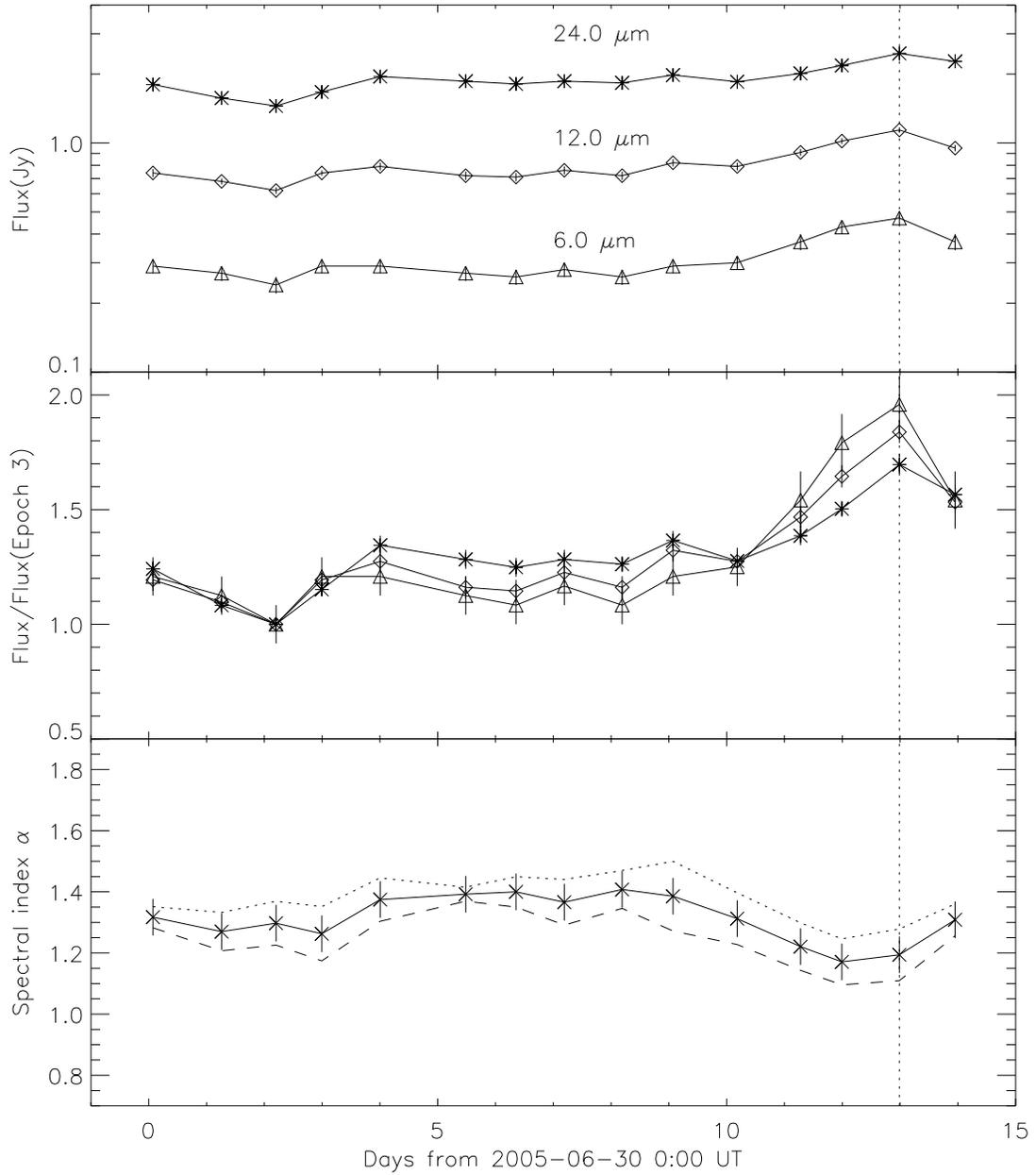}
  \figcaption{{\it Spitzer} IRS mid-IR light curves in three bands for 2005 Jun 30-July 14.
              Top: logarithmic scale; Middle: normalized in each band to the flux at epoch 3. The variability
              amplitude decreases with increasing wavelength at all times. Bottom: spectral index curves, 
              solid line $=$ 6-24 $\mu$m index dotted line $=$ 6-12 $\mu$m index, and dashed line $=$ 12-24 $\mu$m index. 
              The spectrum becomes harder during the bright flare that peaks on 2005 July 13, indicated with a vertical line.}
\end{figure}

\begin{figure}[ht]
  \plotone{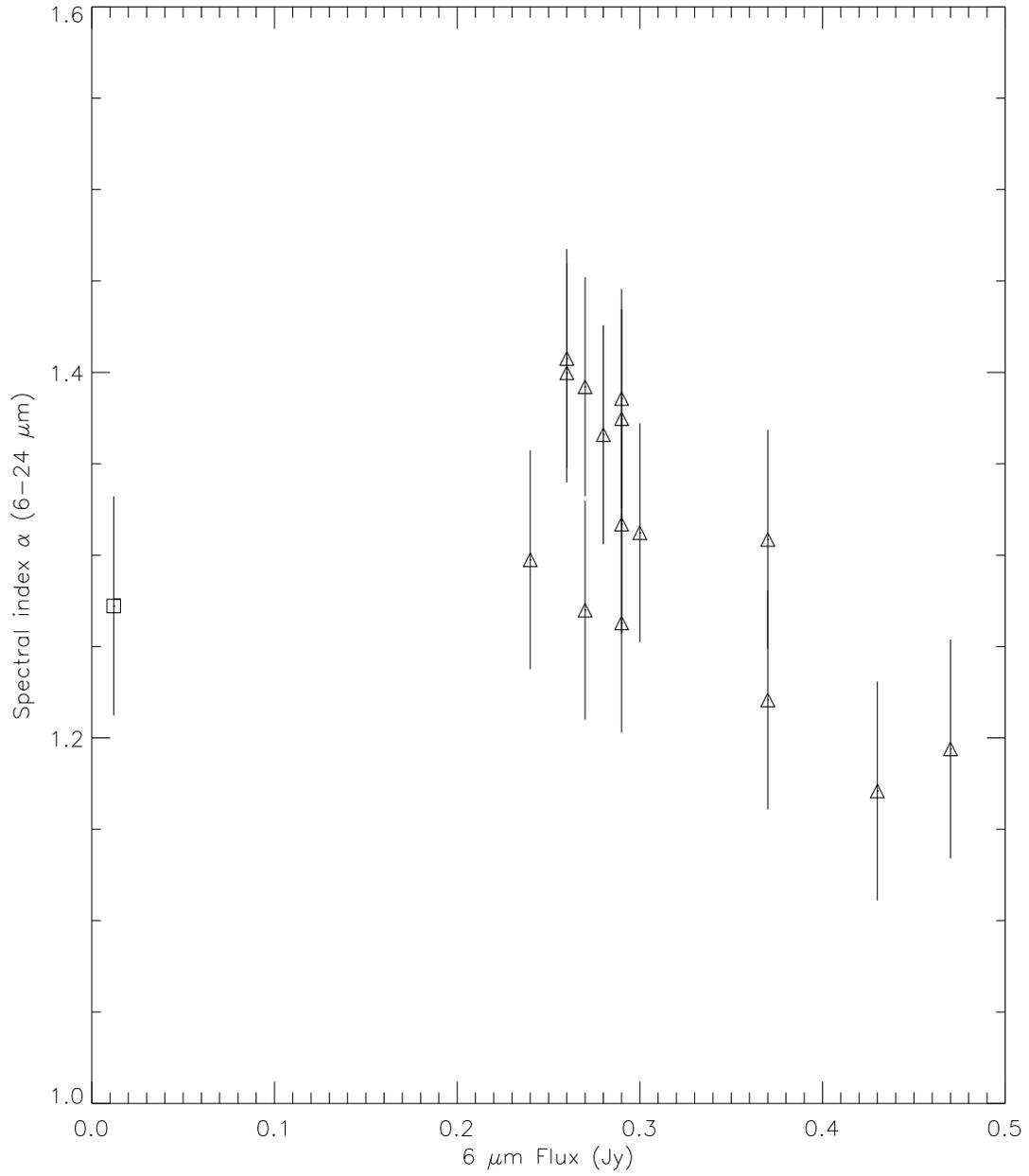}
  \figcaption{Color-flux diagram. The 6-24 $\mu$m spectral index is plotted vs. 6 $\mu$m flux. Triangles: 2005 June-July,
              square: 2006 December. During the 2005 epochs, there is a general trend for harder spectral index
              at greater flux. However, there is not a 1:1 correspondence, likely because the flux and spectral index
              depend on the detailed history of each IR flare.}
\end{figure}

\begin{figure}[ht]
  \plotone{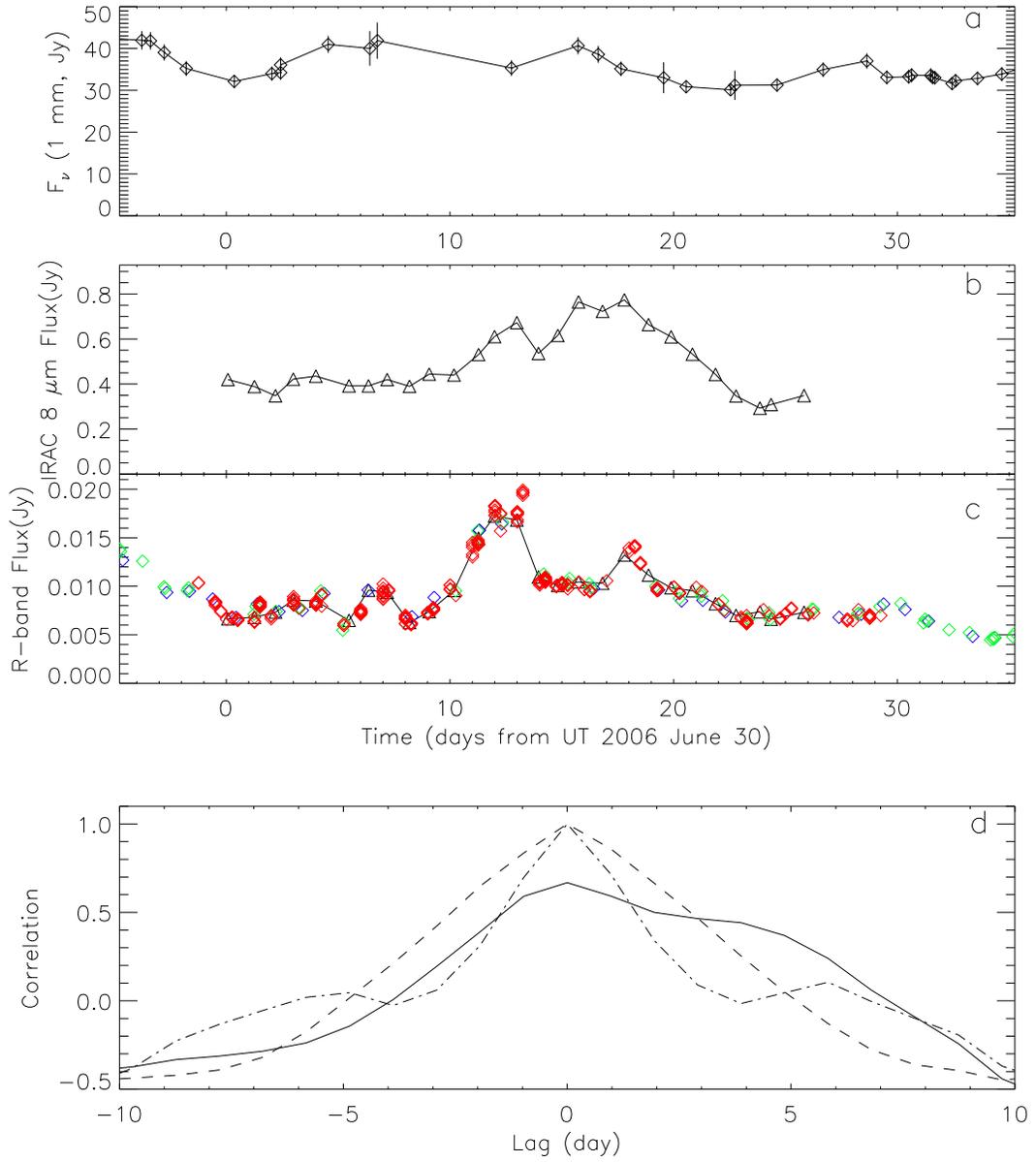}
  \figcaption{Comparison of a) 1.3 mm (230 GHz) radio, b) IRAC 8.0 $\mu$m and  c) R-band light curves. WEBT R-band data \citep{vrb06} 
              are included with the Colgate and Palomar data to improve the sampling during the 2005 June-July Spitzer observation perod. 
              The mm-wave variations do not appear to be correlated with the mid-IR variations, suggesting that emission in the two bands 
              comes from separate regions in the jet. c) Cross-correlation of IRAC 8.0 $\mu$m and R-band light curves (solid line). The 
              autocorrelation curves for IRAC 8.0 $\mu$m (dash), and R-band (dot-dash) are shown for comparison. The IR and optical emission 
              are strongly correlated at zero lag.}
\end{figure}

\begin{figure}[ht]
  \plotone{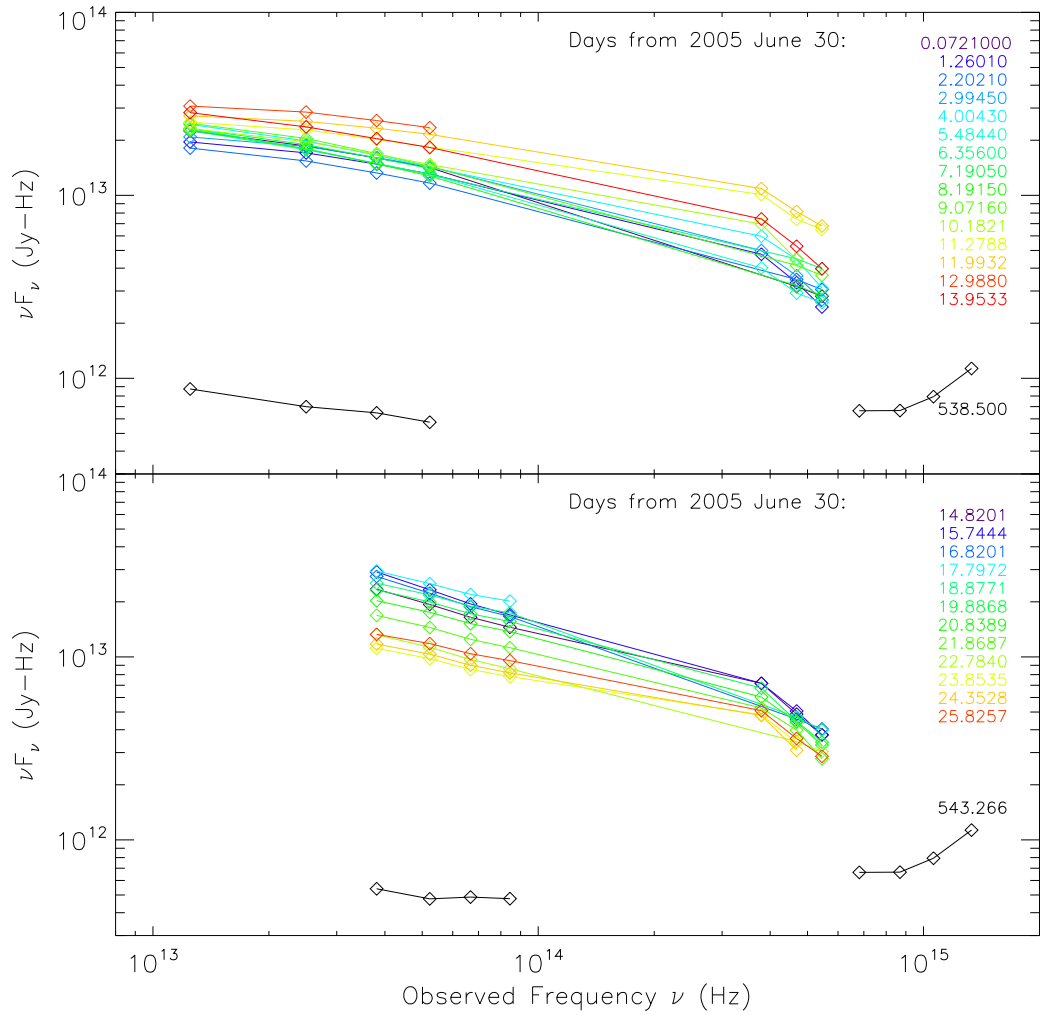}
  \figcaption{SED time series in mid-to-near infrared and optical bands starting on 30 June 2005, and extending for 26 days, 
              broken into two time periods including either the Spitzer IRS data or the  Spitzer IRAC data, with dates distinguished 
              in each frame by colors. Top: Optical-to-infrared (IRS) SED variability during 2005 June-July and on 2006 Dec 20.  
              Bottom: Optical-to-infrared (IRAC) SED variability during 2005 June-July and on 2006 Dec 25. Most of the optical and near 
              infrared measurements in 2005 June-July are from Palomar and Foggy Bottom Observatories; a few points are taken from WEBT 
              data provided by M. Villata \citep{vra07}.  For comparison, we show the rise in the SED (at $\sim 10^{15}$ Hz) attributed 
              by \cite{Rai08} to the Big Blue Bump; B,U, W1, and M2 data points are from the Optical Monitor on XMM, taken on 2006 
              Dec 18-19, (Table 3 of \cite{rvl07} ).}
\end{figure}

\begin{figure}[ht]
  \plotone{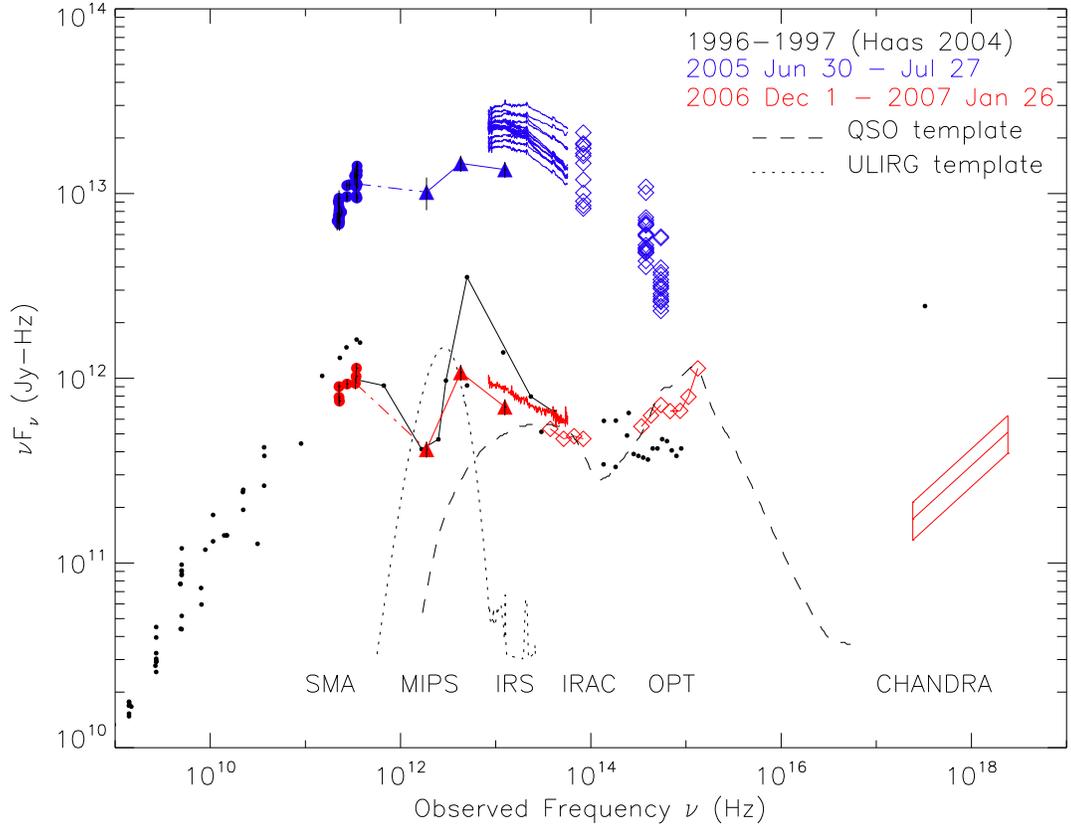}
  \figcaption{3C 454.3 SED. {\it Spitzer} MIPS 24, 70, and 160 $\mu$m photometry (triangles), IRS spectra (solid lines), 
              and IRAC 3.6 $\mu$m and optical photometry (diamonds) are plotted together with other multiwavelength data. 
              The mm and sub-mm points are SMA calibration data. Optical photometry in 2005 is from Foggy Bottom 
              Observatory and the automated Palomar 1.5 m telescope. XMM-OM (UVM2, UVW1, U, B) photometry for 2006 Dec. 
              18-19 are from Raiteri et al. 2007, Table 3. V, R, and I photometry for 2006 Dec 20 are from WEBT data 
              provided by M. Villata \citep {rvl07, vra07}. Both sets have been corrected for extinction. {\it Chandra} 
              X-ray data for the 2007 minimum are are indicated by the error box. The scattered small black dots are 
              radio and optical photometry from the NED database, gathered from 1979-1995. The mean Richards et al. (2006) QSO SED
              (dashed line) and the Rieke et al. $10^{13} L_\odot$ ULIRG template (dotted line) are scaled and overplotted 
              for comparison. Neither of these SEDs is compatible with the shape of the IR bump in the low-state 3C 454.3 
              SED, indicating a dominant nonthermal contribution from the jet. Note the presence of two synchrotron peaks,
              at IR and sub-mm wavelengths, in both the low and high states.} 
\end{figure}

\begin{figure}[ht]
  \plotone{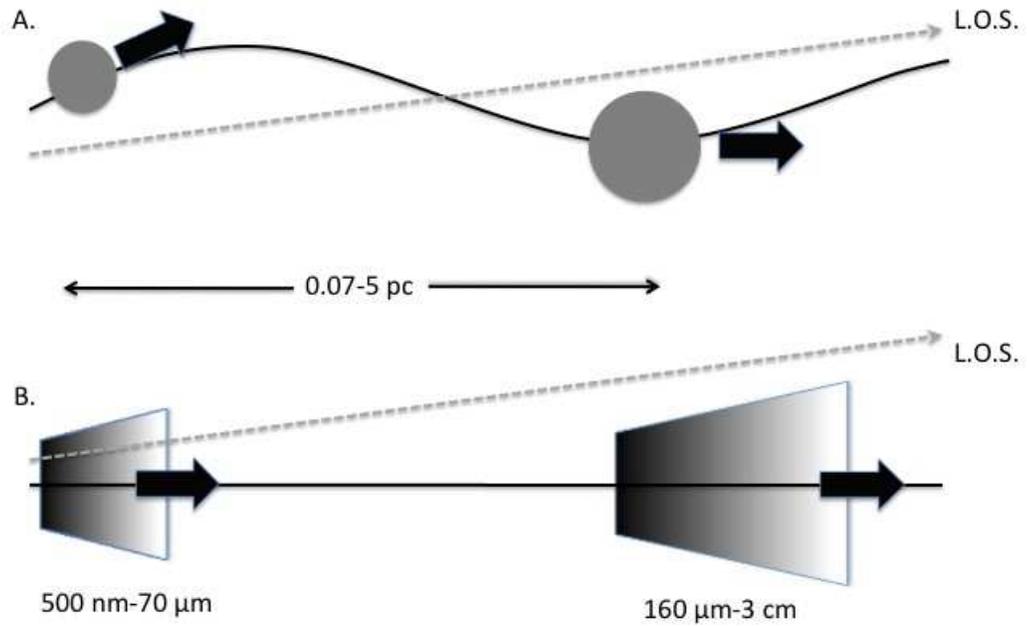}
  \figcaption{A) Helical jet model. The two synchrotron peaks are produced by two relativistic blobs moving at different 
              angles to the line of sight. The IR-optical peak is produced closer to the base of the jet, while the submm peak is 
              produced further out. B) Inhomogeneous jet model. Two radially separated synchrotron emission regions 
              give rise to the IR and submm synchrotron emission peaks.  The IR emission region comes from the base of the jet, 
              while the sub-mm emission region comes from a shock (possibly a jet recollimation shock) at a radius
              of $\sim 0.07-5$ pc. The direction to the observer's line of sight is marked ``L. O. S.''}
\end{figure}

\end{document}